# Recent Progress in Cartilage Lubrication

Weifeng Lin and Jacob Klein*

Department of Materials and Interfaces, Weizmann Institute of Science, 76100 Rehovot, Israel

*E-mail: Jacob.klein@weizmann.ac.il (J.K.).

**Abstract:**

Healthy articular cartilage, covering the ends of bones in major joints such as hips and knees, presents the most efficiently-lubricated surface known in nature, with friction coefficients as low as 0.001 up to physiologically high pressures. Such low friction is indeed essential for its well being. It minimizes wear-and-tear and hence the cartilage degradation associated with osteoarthritis, the most common joint disease, and, by reducing shear stress on the mechanotransductive, cartilage-embedded chondrocytes (the only cell type in the cartilage), it regulates their function to maintain homeostasis. Understanding the origins of such low friction of the articular cartilage, therefore, is of major importance in order to alleviate disease symptoms, and slow or even reverse its breakdown. This progress report considers the relation between frictional behavior and the cellular mechanical environment in the cartilage, then reviews the mechanism of lubrication in the joints, in particular focusing on boundary lubrication. Following recent advances based on hydration lubrication, a proposed synergy between different molecular components of the synovial joints, acting together in enabling the low friction, has been proposed. Additionally, recent development of natural and of bio-inspired lubricants is reviewed.

# 1. Introduction

Friction between two surfaces is often discussed within the context of Amontons' law $F = \mu N$, where a coefficient of friction $\mu = F/N =$ (force to slide the surfaces)/(load compressing the surfaces). Amontons' law counterintuitively indicates that friction force is directly proportional to the loading, independent of the apparent contact area and sliding speed,[1, 2] though the presence of lubricants or soft boundary layers may modify this simple law.[3, 4] The essence of friction is considered to be an energy dissipation process when two surfaces slide past each other,[5] as discussed further below. The energy lost in friction may be in the form of heat, sound, and other forms of energy, while lubrication is the process by which such energy lost is minimized.[6] Frictional energy dissipation is conveniently considered in two main categories. Viscous dissipation occurs when two surfaces separated by a thin liquid film slide past each other. The resistance to motion arises as the viscous layer is sheared, and the friction can be written as $F = A \cdot \eta_{\text{eff}} \cdot \frac{dx}{dt}/D$, where $A$ is the surface area, $\eta_{\text{eff}}$ is the effective viscosity (which may be shear-rate dependent), $dx/dt$ is the relative sliding velocity between the two surfaces past each other, and $D$ is the thickness of the fluid film, assuming the usual stick boundary conditions at the surfaces. Therefore, energy dissipation $E_{\text{viscous}}$ can be represented as $E_{\text{viscous}} = A \cdot \int [\eta_{\text{eff}} \cdot \frac{dx}{dt}/D] \cdot dx$ over the sliding distance (as shown in **Figure 1**A). Boundary friction is the other major mode of frictional energy loss, which arises when the two surfaces, or boundary layers attached to them, are in molecular contact during sliding. Energy dissipation may then arise, for example, from irreversibly adhesive bonds breaking and reforming, such as van der Waals or opposite charge-charge adhesive interactions between the contacting molecules;[7-9] or irreversible processes arising from overcoming energy barriers as the surfaces slide past each other;[10-13] or from dis-entanglement of polymer molecules attached to the opposing surfaces;[14-17] and so on.[18] In contrast to viscous dissipation, boundary friction is more complicated to describe theoretically, but generally has a much weaker sliding-velocity-dependence than the linear velocity-dependence of viscous energy dissipation.[19] As boundary friction occurs at the slip plane between the boundary layers rather than the underlying substrates, such friction is largely independent of the underlying substrates (which are not themselves in contact) and depends mostly on the physicochemical properties of the contacting boundary layers (as shown in **Figure 1**B). Energy dissipation modes other than boundary friction, such as viscoelastic losses in the near-surface-region, will depend on the mechanical properties of the substrate, and may also contribute to the overall sliding friction. In the context of modelling and understanding biological lubrication, which is a focus of this progress report, this substrate-independence of boundary friction is an important characteristic of the boundary layers. This is because it implies that experimentally measured friction between model substrates coated with given boundary layers may provide insight into the boundary friction between living biological tissue, such as articular cartilage, so long as it is coated with similar boundary layers *in vivo*. This is schematically illustrated in **Figure 1**B.

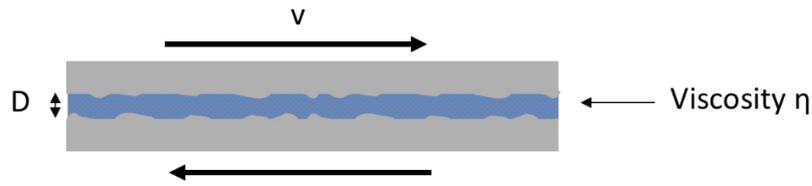

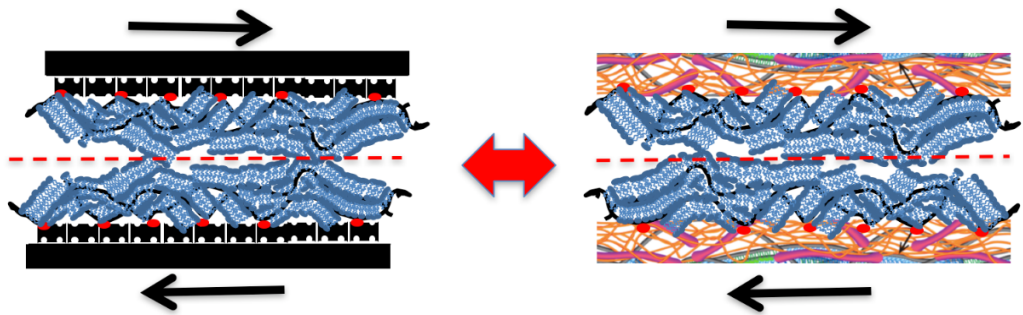

**Figure 1.** Two main modes of lubrication in cartilage. (A) In fluid film lubrication, shear stress (shear or friction force per unit surface area) σ, $\sigma = \eta v/D$, where $D$ is the thickness of a uniform film, $\eta$ is the viscosity of film, and $v$ is the sliding velocity, which indicates that shear stress varies linearly with sliding velocity. (B) In boundary lubrication, the frictional dissipation takes place at the slip plane indicated, and is largely independent of the different substrates shown.

Functioning of mammalian bodies, humans not least, involves extensive sliding of tissues and organs past each other, as in synovial joints, tendons gliding within their sheath, eyelids past the cornea, internal organs in the human body and so on, which often exhibit ultralow sliding friction as part of their healthy function. The major human synovial joints such as hips and knees are prime examples. Articulation occurs with sliding friction coefficients as low as $\mu \approx 0.001$[20] up to physiologically high pressures[21] (withstanding loads, due to additional ligament and musculature stresses, up to several times the body weight[22]) and shear rates over a wide range[23] (from rest to ca. $10^5$–$10^6$ sec$^{-1}$). The articular cartilage coating the joints and enabling this thus presents the most remarkably lubricated surface in nature, unmatched by even the most advanced synthetic materials. Such lubrication is essential to healthy function of the joints, and its breakdown is associated with joint dysfunctions such as osteoarthritis (OA), a chronic musculoskeletal condition. OA is thought to result from a cascade of events, beginning with damage to the articular cartilage, often initiated by some trauna (or by old age),[24] leading to a disruption of lubrication. The resulting increase in friction at the articular cartilage surface in turn causes progressive structural and functional damage of the cartilage tissue and the subchondral bone, ultimately leading to joint failure with pain and disability.[25] OA, as a common, debilitating disease, is a leading cause of chronic pain and disability in adults, affecting many tens of millions in Europe and the US alone.[26] Moreover, damaged cartilage is not capable of growing or healing itself due to its lack of blood vessels and low density of chondrocytes.[27] OA is generally not curable or reversible (though some intriguing evidence of cartilage regrowth has been reported following joint dis-traction treatments[28]), but treatments (such as hyaluronic acid (HA) injections, cyclooxygenase 2 inhibitors, non-steroidal anti-inflammatory drugs (NSAIDs), and joint replacement surgery, etc.) have been used to alleviate joint pain and improve joint mobility.[29] Most of the non-surgical treatments may have unwanted side effects (such as severe pain at the injection site after intra-articular injection of HA, or NSAIDs-associated adverse gastrointestinal effects) but only short-term pain relief, thereby hampering their long-term use.[30] Thus, it is important to better understand the molecular-level origins of such low energy dissipation on sliding in natural joints. The answer to this would not only provide new insight, but could point to new treatment approaches to alleviate pain or even reverse degenerative processes.

In this progress report we present a detailed overview of the advances of the past decade or so in understanding the origins of efficient boundary-lubrication in the body. We are especially interested in boundary lubrication processes, and particularly (though by no means exclusively) with reference to cartilage and synovial joints, as these operate in the most mechanically-stressed conditions in the body. In the following section, we present the structure and constituents of articular cartilage, and explore the intriguing connection between bio-lubrication and anabolic/catabolic cellular responses arising from mechanotransductively-driven gene-regulation. In subsequent sections we discuss the mechanisms of articular cartilage lubrication, and following that we describe hydration lubrication, and several systems where the hydration lubrication effect has been observed and studied. We then discuss the molecular synergy relating to boundary lubrication which arises when different types of molecules act together as boundary layers at articular cartilage surfaces. Finally we review some synthetic bio-lubricants that have been inspired by biological systems. We conclude this progress report by considering the development of new materials and approaches based on the developing molecular-level understanding of biological lubrication processes, and the prospects of such approaches for a variety of advanced applications and therapies.

## 2. Structure and Constituents of Articular Cartilage

### 2.1. Cartilage Structure and Composition

In general, there are three different distinct cartilage types in the body based on biochemical properties: hyaline (e.g. articular cartilage), elastic (e.g. ear cartilage) and fibrous (e.g. intervertebral disc) cartilage.[31] The articulating end of joint-bones (**Figure 2**A) is coated by articular cartilage with a thickness of 1~4 mm,[32, 33] which transmits joint load while keeping a low frictional coefficient. Articular cartilage is a highly specialized and structured connective tissue composed of a sparse distribution of chondrocytes (<2% to 5% by volume of articular cartilage[36]) surrounded by a dense extracellular matrix (ECM). Water (either bound, i.e. water of hydration, or unbound water) is the most abundant constituent (ca. 70% of the wet weight) in the healthy articular cartilage. Water content increases from ca. 65% near the subchondral bone to ca. 80% at the outer superficial zone,[37, 38] so that we may think of cartilage as a complex hydrogel. Due to the lack of blood vessels, water within the cartilage plays a key role in transporting nutrients to the chondrocytes,[39] which are the only cell type in the tissue.

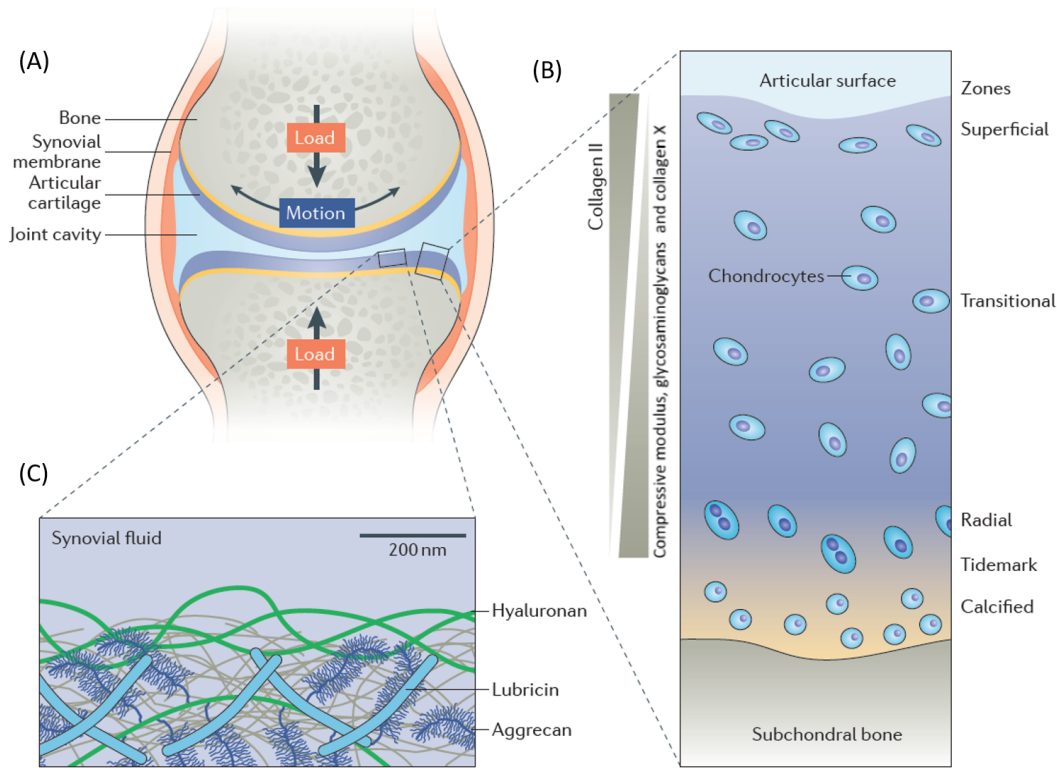

**Figure 2.** Structure and constituents of articular cartilage. (A) A synovial joint, enclosed within the synovial membrane which is lined with specialized cells (synoviocytes) and contains the synovial fluid within the joint cavity. (B) The articular cartilage layer is organized into four main zones located above the subchondral bone. Chondrocytes (cartilage cells) occupy less than 5% (volume fraction) of cartilage tissue. The matrix composition of collagen II, collagen X, and glycosaminoglycans (GAGs) of the tissue, as well as its modulus has a depth-dependence as indicated. (C) Structure of the most superficial layer of articular cartilage. Linear hyaluronic acid (HA, green), bottle-brush like aggrecan (violet), and lubricin (light blue) are major macromolecules involved in cartilage lubrication, with lubricin reported both at the surface of cartilage and at the outer superficial zone, while phospholipids are also present on the cartilage surface but are not shown here. Panel B reproduced with permission from ref. [34]. Copyright 2012, John Wiley and Sons, Inc. Panel C reproduced with permission from ref. [35]. Copyright 2002, AAAS.

Other than water, cartilage is comprised of a large number of different proteins, macromolecules and lipids;[40, 41] here we describe the main ones thought to be most relevant to its mechanical and frictional properties.[20, 42] Collagens (55~75%, mainly, and especially near the cartilage outer surface, type II collagen),[43] proteoglycans (15~30%),[44-46] and lipids (mostly phospholipids (PLs), ca. 10%)[47, 48] make up most of the dry weight of cartilage. Collagens, the most abundant structural macromolecule in cartilage, provide tensile stiffness and strength to the cartilage.[49, 50] Collagen type II is a fibrillar protein, accounting for ca. 90% of the collagen in articular cartilage, forming fibrils and interacts with other collagens, proteoglycans, and non-collagenous proteins.[51] Classifying the cartilage by variations in its structure and composition, three zones (**Figure 2**B) – the deep/radial zone, the middle/transitional zone, and the superficial zone – may be distinguished, from the bottom to the surface of articular cartilage.[52] Collagen content is highest at the articular surface and decreases with depth, as indicated in **Figure 2**B.[43]

Large-diameter collagen fibrils are arranged perpendicular to the tidemark, which is at the border between the deep zone (comprising ca. 30% of articular cartilage volume) and the calcified cartilage which borders the subchondral bone itself (**Figure 2**B). The middle/transitional zone provides an anatomic and functional bridge between the deep and superficial zones. It represents 40–60% of the total cartilage volume, where collagen fibres are organized obliquely, and the chondrocytes are more spherical and sparsely distributed. In the superficial zone, the collagen fibers (mainly type II) are aligned parallel to the articular surface, and are thereby better able to resist compressive, tensile and shear forces. On the other hand, the concentration of proteoglycans is very low in the superficial zone, but increases with cartilage depth, and with it the compressive modulus of the tissue (**Figure 2**B).[53, 54]

Proteoglycans are composed of a core protein with covalently attached glycosaminoglycan (GAG) chains (**Figure 2**C). The side-chains are linear carbohydrate polymers carrying a high density of negative charges under physiological conditions due to $-SO_3^-$ or $-COO^-$ groups.[55] Aggrecan is the most abundant macromolecule (ca. 8%) present within the collagen fibril network. It is a high molecular weight proteoglycan, composed of a protein core and negatively charged GAG side-chains (ca. 20–40 nm long) of chondroitin sulfate and keratan sulfate covalently attached to it, forming a bottle-brush-like structure (**Figure 2**C). Aggrecans are in turn non-covalently attached by one end to HA, a long, linear highly-charged polysaccharide, via link-proteins,[56] as shown schematically in **Figure 2**C. HA, lubricin, and PLs are the main components in the superficial zone adjoining the cartilage surface, and are also present in the synovial fluid. Lubricin, a mucinous glycoprotein which is highly conserved (thus suggesting it plays an important role in the evolutionary development of joints), is another important proteoglycan that is present on the superficial layer of articular cartilage, and plays a major role in cartilage integrity and synovial homeostasis.[57-59] Phosphatidylcholines (PCs), that is PLs with two hydrophobic acyl tails attached to a highly-hydrated zwitterionic phosphocholine headgroup, form the largest group of PLs (ca. 40%) in cartilage and in synovial fluid, which is the viscous liquid bathing the cartilage within the joint cavity surrounded by the synovial membrane (**Figure 2**A).[47, 60] PCs are also the major component of biological membranes.

## 2.2. Chondrocytes

Chondrocytes (the only cell type in cartilage) derive from mesenchymal stem cells, and are responsible for maintaining the surrounding cartilage matrix. They do this by sensing changes in the composition of the ECM arising from degradation of macromolecules and demands at the interface during injury or degeneration. Chondrocytes respond to such changes through appropriate synthesis of the required molecules, such as type II collagen, proteoglycan, and other specific non-collagenous proteins.[61] The balance between matrix synthesis and degradation is controlled by the relative amounts of growth factors and cytokines in the cartilage or synovial fluid. The behavior of chondrocytes may be highly regulated by load and pressure, and the way in which these vary dynamically, though the detailed mechanisms, and why they appear to respond differently to shear stresses as opposed to compressive loading, are not yet fully understood.[62-64] In particular, recent studies suggest that joint tissues are highly mechanosensitive, and friction may play a subtle role through its effect on chondrocytes.[65] The reason is that higher friction at the interface of cartilage during articulation results in larger frictional shear stresses which can transmit through the extracellular matrix to the embedded chondrocytes. Both compressive and shear stresses on chondrocytes are known to regulate their gene expression.[64-67] However, while compressive stresses (mechanical loads dependent on body weight and muscle tension) are undoubtedly important for the maintenance of cartilage (anabolic processes),[68] shear stresses may lead to upregulation of catabolic enzymes (catabolic processes). These include aggrecanase,[69] hyaluronidase,[70] phospholipases,[71] and matrix metalloproteinases (MMPs),[72] known to involved in matrix breakdown. Shear stresses may also be associated with lower expression of cartilage-specific markers, such as type II collagen and aggrecan.[73] Meanwhile, chondrocytes rely mostly on the circulation of synovial fluid, and diffusion into the cartilage, to provide nutrients for survival.[74] Walking, jogging, and other normal activities are good ways for inducing cycles of compression and decompression, which results in the circulation of nutrients to the interior of the cartilage tissue. Conversely, loss of mobility due to OA can seriously affect the health of chondrocytes, which also decreases the anabolic gene expression that maintains the cartilage, in a detrimental cycle.[75]

This suggests the following scenario correlating friction with OA (**Figure 3**). Friction that is higher than that associated with healthy, well-lubricated cartilage, which may arise from age-related factors, or some initiating event such as accident or sport traumas,[76] leads to increased shear stress $\sigma_s = \mu P$ at the cartilage surface, where $\mu$ is the coefficient of friction and $P$ the normal contact stress. This increased stress results in a higher shear strain $\varepsilon_s$ transmitted via the cartilage, with shear modulus $K_s$ to the chondrocyte cells embedded within the tissue in the near surface region, so that $\varepsilon_s = \sigma_s/K_s = \mu P/K_s$. This increased shear strain in turn causes upregulation of (catabolic) cartilage-degrading enzymes via cell mechano-transduction.[77] The cartilage surface degraded by such increased enzyme production is associated with higher-friction sliding motion as the joints articulate, for example because it becomes rougher, or because its residual lubricating boundary layer is less efficient than prior to its degradation. This higher friction then leads to higher shear strains on subsequent sliding and thereby yet more catabolic enzyme upregulation, and so on in a self-reinforcing cycle, resulting in progressive damage to the cartilage and eventually its total destruction, as indicated in **Figure 3**. A more detailed description is given in **Figure 3**. It is clear that since $K_s$ and $P$ are given parameters related to cartilage bulk structure and to body weight and musculature stress on the joint, the main way to decrease the shear strain and thus catabolic regulation of the chondrocytes is through reduction of the friction coefficient $\mu$.

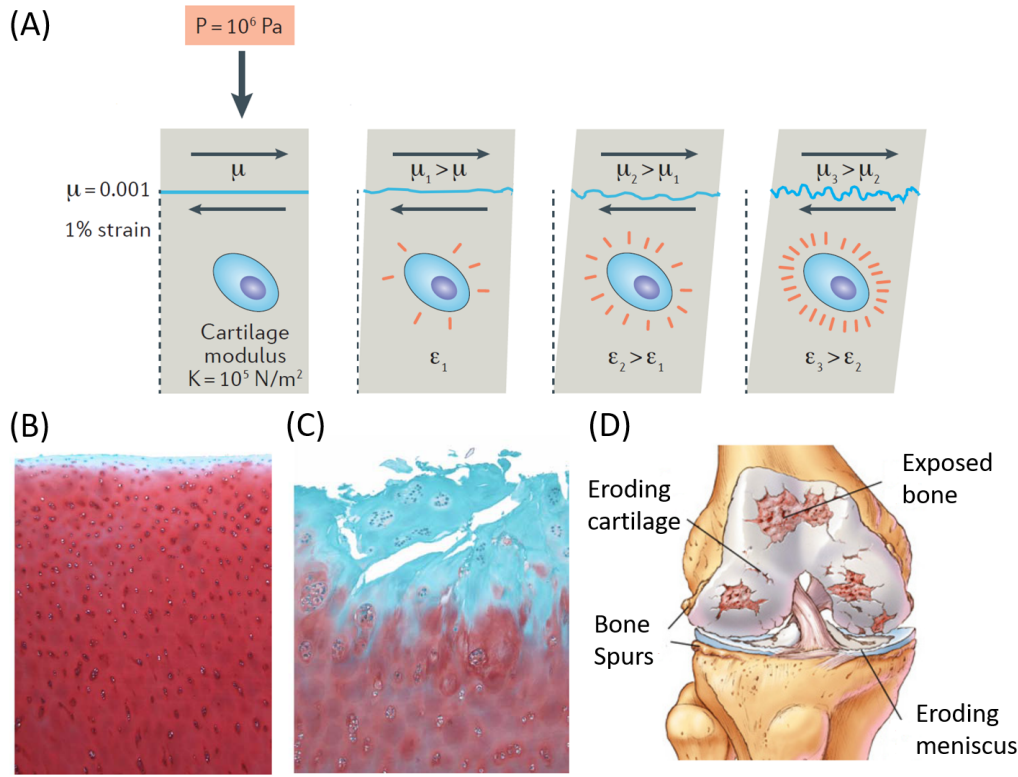

**Figure 3.** (A) For surfaces compressed by a load $F_n$ over a contact area $A$, the mean pressure is $P = F_n/A$. When the surfaces slide past each other the friction force $F_s$ is equivalent to a shear stress $\sigma_s = (F_s/A)$ which causes a shear strain in the material given by $\varepsilon = (\sigma_s/K_s)$, where $K_s$ is the (effective) shear modulus of the material; the dashed line is the unstrained state. Thus, given the friction coefficient $\mu = (F_s/F_n)$, it follows that the shear strain $\varepsilon_s = (\mu P/K_s)$. For articular cartilage layers sliding past each other, the strain arising from the friction will be transmitted through the tissue to any embedded chondrocytes in the vicinity of the surface (typically to depth of order a hundred microns from the surface, and bearing in mind that $K_s$ and thus $\varepsilon_s$ vary with depth from the surface. For characteristic values of $K$ (say ~$10^5$ N/m²),[54] ($\mu = 0.001$ say) and $P = 10^6$ N/m² ($\approx$ 10 atm, characteristic of mean physiological pressures[21, 78, 79] in human articular cartilage which vary between 0 and several 10's atm), we find the shear strain $\varepsilon = (\mu P/K_s) \approx 1\%$ (though as $K_s$ increases with depth, the strain will correspondingly decrease). Comparable or somewhat higher strains are known to readily elicit significant mechanotransductive responses from chondrocytes.[62] From left to right, we illustrate schematically the proposed self-reinforcing cycle of cartilage

degradation described in the text, arising as excessive friction ⇒ excessive shear strain on chondrocytes ⇒ pro-catabolic chondrocyte gene regulation via mechanotransduction (i.e. expression of cartilage-degrading enzymes, indicated schematically as greater density of lines emanating from cells) ⇒ cartilage degradation (shown schematically as rougher surfaces) ⇒ higher friction and shear strain and so on. We have assumed in this schematic (by extrapolating from earlier results[62, 64, 67]) that expression of cartilage-degrading enzymes may increase with the strain. (B) A histology image of healthy articular cartilage shows the cartilage surface is smooth. (C) A histology image of osteoarthritic cartilage exhibits cartilage matrix loss extends from the superficial zone into deeper zones. (D) This schematic shows degenerated cartilage in osteoarthritis. Reproduced with permission from ref. [80]. Copyright 2018, American Institute of Physics Publishing.

## 3. Models of Lubrication at Cartilage Surfaces

Friction coefficient for normal whole synovial joints have been reported in the range $\mu \approx$ 0.002–0.02 (typical values are shown in **Table 1**).[23, 42, 81-87] However, measurements of friction on living articular cartilage surfaces, or even on whole joints, for example by using a joint as the fulcrum of a pendulum,[88, 89] are likely to be affected by other energy dissipation mechanisms that occur during articulation. These typically arise from viscoelastic distortion of adjacent tissues (ligaments, muscles, synovial membrane) as the joint rotates.[90] Such viscoelastic losses are not related to cartilage-cartilage friction *per se*,[91] but would manifest themselves as measured friction forces. Thus it is very likely that, low as the values appearing in **Table 1** are, the intrinsic, cartilage-against-cartilage value of the friction coeeficient in living joints is even lower.

Why is friction in a healthy synovial joint so low? Despite many decades of study,[42, 92-96] the precise lubrication mechanism remains elusive. The complex structure and composition of articular cartilage surfaces makes it likely that multiple dissipation pathways (for example fluid-film viscous losses, boundary friction, and viscoelastic losses within the near-surface cartilage tissue) are responsible for the low friction and wear of articular cartilage (**Figure** 1).

Table 1. Friction coefficient for articular cartilage.

| Substrate | Lubricant | μ | Comments | Ref. |
|---|---|---|---|---|
| **Horse knee joint** | Saline | 0.02 | Pressure ≈ 10 atm | [82] |
|  | Synovial fluid | 0.02 |  |  |
| **Dog ankle joint** | Buffered saline | 0.0099 | Pressure ≈ 14 atm; | [83] |
|  | Synovial fluid | 0.0044 |  |  |
| **Human hip joint** | Ringer's buffer | 0.013 | Pressure ≈ 10–20 atm | [97] |
|  | Synovial fluid | 0.008 |  |  |
| **Bovine knee joint** | Veronate buffer | 0.0052 | Pressure ≈ 50 atm | [87] |
|  | Synovial fluid | 0.0025 |  |  |
| **Bovine knee joint** | Veronate buffer | 0.011 | Pressure ≈ 80 atm | [86] |
|  | Synovial fluid | 0.011 |  |  |

## 3.1. Fluid Film Lubrication

Fluid film lubrication is a class of lubrication mechanisms having in common that a fluid film completely separates sliding surfaces (a film considerably thicker than the surface asperities). The earliest fluid film models,[98] following the classic theory of hydrodynamic lubrication in machines,[99] were later extended to account for the deformational behavior of articular cartilage during motion.[100] Other models, pioneered by McCutchen,[101, 102] Ateshian,[103, 104] and others,[105-107] consider films between the articulating surfaces that arise as a result of exudation of the interstitial fluid on compression of the cartilage. In these the fluid film prevents direct contact between surfaces and minimizes friction and wear **(Figure 3**A), while the interstitial fluid pressure supports significant portions of the load on the cartilage. A corollary of this model – whether or not a fluid film actually intervenes between the sliding surfaces – is that the actual contact stress between the cartilage surfaces may be substantially lower than the high pressures measured directly at the cartilage interface.[21, 78, 79] This is because much of the load is born by the interstitial fluid pressurization.[103] For this reason, many of the studies of friction on cartilage explants were carried out at physiologically-relevant strains but at low pressures (up to ca. 2 atm), which is about a hundred-fold lower than the measured maximal measured pressures at articular cartilage surfaces in whole joints[86, 87] (whether these measured pressures are due to interstitial fluid pressure or to actual boundary contact pressures). These ideas were quantified by considering the tissue as an incompressible porous-permeable solid matrix (the collagen-proteoglycan network),[55, 108] while the interstitial fluid is modeled as an incompressible fluid. These 'biphasic' scenarios[108] and their extensions were used to model the response of articular cartilage to loading,[109] and have been examined in studies using excised cartilage samples.[85, 110-113] Krishnan et al.[85] performed an important study testing the interstitial fluid pressurization model carried out at ca. 1.5 atm applied pressure (on a 6-mm diameter explant), showing an increase in the friction coefficients from ~ 0.04 to 0.18 within loading period of 15 mins (and from ~ 0.04 to 0.1 in about 100 sec). This large and rapid increase in friction could be closely correlated with the simultaneously measured reduction in interstitial fluid pressurization with load time, and thus presented as a demonstration of the validity of the model.[114, 115] Interstitial pressurization that bears much of the load between opposing cartilage surfaces *in vivo* has been a central working hypothesis of many previous studies;[116-120] in contrast, the focus of this review is on progress and advances in our understanding of cartilage boundary lubrication. That is, the friction arising from energy dissipation when the sliding, opposing surfaces are in molecular contact. Such boundary friction is the main mechanism of frictional dissipation between two compressed cartilage surfaces as they slide past each other.

## 3.2. Boundary Lubrication

At high pressures between the cartilage surfaces, any fluid film between them may be squeezed out so that either partial or full molecular contact between the surfaces is established. A simple estimate, based on Hertzian contact mechanics and the known modulus of the cartilage superficial zone, suggests that affine molecular contact is readily formed at the typical pressures acting between the cartilage surfaces.[42] Indeed, this is estimated to happen at contact stresses higher than local contact stress $P_a \approx (K/\pi)$, where $K \approx 10^4$–$10^5$ N/m$^2$ is the near surface Young's modulus of cartilage.[54] That is, the near-surface articular cartilage layer is sufficiently compliant so that under such typical contact stresses ($> P_a$) between cartilage surfaces (from sub-MPa and higher) any bumps and roughnesses on each surface will be distorted to form intimate contact with the opposing surface.

For the case of only partial molecular contact, one expects that both fluid film and boundary lubrication will be active.[23, 92, 99, 121] Thus, the variation of the sliding friction with sliding velocity in whole joints has been measured both as independent of velocity, supporting a boundary mechanism (a conclusion reached by several other direct studies on whole joints[91, 97]) and as decreasing with velocity, suggesting a mixed regime.[92] It is especially noteworthy that there is a very non-uniform lateral pressure distribution across the cartilage surface measured in whole joints, both *in vivo* using direct pressure sensors, and in cadavers using pressure sensitive films.[21, 78, 79] Such large, persistent lateral-non-uniformity of the pressure, is at odds with the model whereby interstitial pressurization in the cartilage bears much of the load between the surfaces, as that would suggest a uniform stress across the contact area. It indicates rather that substantial regions of the cartilage are in molecular boundary contact at high pressures, while separated by intervening regions that may reflect the presence of fluid films at lower pressures. A recent study[122] found that lipid-based lubricants, attached by binder molecules to form a boundary layer on an intrasynovial tendon surface, lead to a striking and sustained reduction in tendon/sheath friction ($\mu \approx$ 0.02–0.03, albeit at quite low contact stresses), which remains constant over at least 500 back-and-forth cycles. This compares with significantly larger, and increasing, friction in the absence of such lubricants, indicating strongly that, for this model system at least (which has structural and compositional similarities with articular cartilage), the dominant friction reduction is due to boundary lubrication. It is important to bear in mind that any viscous losses due to shear of fluid films between sliding cartilage surfaces are very much lower than frictional dissipation processes between molecularly-contacting surface-boundary layers. As an example, if a uniform film of thickness as low as say 0.1–1 μm, comparable to the roughness of the cartilage surface[42, 123] and viscosity similar to water[124] separates two cartilage surfaces sliding past each other at 1 cm/sec, under a pressure of 1 MPa, the frictional drag arising from viscous losses in the film corresponds to a friction coefficient of only ca. $10^{-4}$–$10^{-5}$, much lower than measured friction coefficients for cartilage (**Table 1**). For this reason, as noted above, boundary friction [79, 91, 112] is the dominant dissipation mode when articular cartilage surfaces slide past each other. Given the importance to joint health[125-127] of low cartilage friction, as decribed above (and illustrated in **Figure 3**), there have therefore been comprehensive efforts to understand and identify the characteristic structure and molecular composition of the boundary layers at the cartilage surface.[41, 128-132]

In a number of these studies whole joints or excised cartilage surfaces (cartilage explants) were used,[86, 91, 97, 113, 117, 120, 133-135] while others involved model substrates, which were coated with candidate molecular boundary layers.[41, 136-142] Both approaches have their advantages and disadvantages. Thus, the former more clearly resemble cartilage in their bulk properties, though it is difficult to control their precise surface boundary composition and roughness. More importantly, the nature of the cartilage surface is known to change significantly once it is manipulated or removed from its unperturbed *in vivo* environment,[143-148] and its micro-mechanical characteristics also change.[129] Such changes imply that measurements on cartilage explants may not be reflecting the surface interaction properties, including boundary friction, of *in vivo* cartilage surfaces in the unperturbed joint.

Investigations using model substrates employ a number of techniques to measure the boundary friction. These include colloidal probe scanning techniques,[149, 150] where a colloidal particle attached to the cantilever of an atomic force microscope (instead of a cantilever with a sharp tip) slides past molecular layers on substrates such as glass, and surface force balance (SFB) (or equivalently surface force apparatus (SFA)) methods.[151, 152] SFBs, which have been used widely in our group and by others, can measure directly the normal and frictional surface forces between boundary layers of different molecules, attached to two essentially parallel, opposing, atomically-smooth mica surfaces. A schematic of the SFB is inset in **Figure 4**. This technique provides state-of-art sensitivity and resolution in measuring of normal and shear stresses between interacting surfaces, as well as the ability to measure the absolute thickness of the surface attached boundary layers to within 0.2–0.3 nm. Model substrates – such as silica surfaces or molecularly-smooth mica – are clearly different to articular cartilage, but they allow close control over surface roughness and the detailed nature of the boundary layers. For this reason, they may provide insight at the molecular level into the mechanisms of friction which occur at the slip plane (boundary lubrication), and which are largely independent of the substrate (**Figure 1**B).

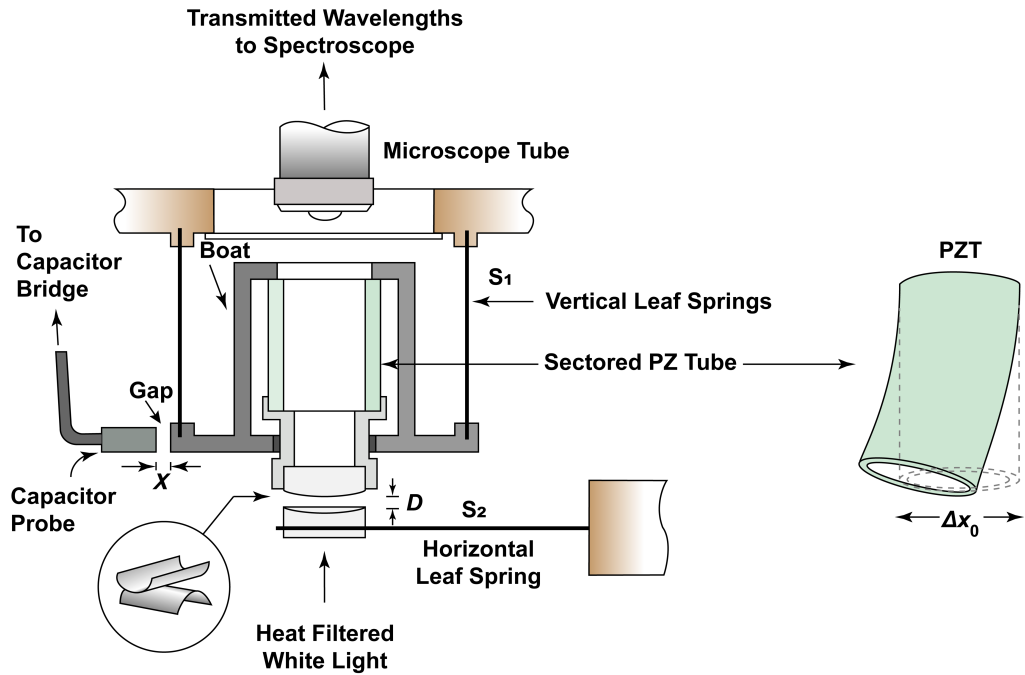

**Figure 4.** A realistic schematic of the SFB used to measure directly normal and shear forces between molecularly smooth mica surfaces.[153, 154] Heat-filtered white light undergoes multiple beam interference in passing through the half-silvered mica sheets (in a crossed cylindrical configuration, inset) to yield absolute separation $D$ between them to within 0.2 nm resolution. A sectored piezoelectric tube (PZT, shown enlarged on the right and illustrating the lateral motion when opposing sectors are made to expand/contract by equal and opposite voltages) is mounted via a rigid stainless-steel boat onto leaf springs $S_1$. Shear forces are measured via the bending of $S_1$, monitored by the changes in the thickness $x$ of a small gap between a fixed capacitance probe and the moving boat. Normal forces are measured via the bending of $S_2$, monitored by the change in wavelength of fringes of equal chromatic order. Reproduced with permission from ref. [154]. Copyright 1998, American Institute of Physics.

Several groups have studied the boundary lubrication properties of different molecular species (as shown in **Table 2**) which are believed to be implicated in cartilage boundary lubrication, or resembling them. The main molecules believed to be associated in cartilage boundary lubricating layers, either alone or in combination, are HA, aggrecans, lubricin, and PLs, as well as others which are briefly considered below. HA was early on considered the molecule responsible for the lubrication of the cartilage surface,[133] as being a long linear polymer, it is the molecule largely responsible for the high viscosity at low-shear rates of synovial fluid, which was considered to be the cartilage lubricant. Subsequently it was demonstrated that the enzymatic removal of HA made little difference to the lubrication ability of synovial fluid.[87] HA has often been, and is still, used as an intra-articularly injected viscosupplementation[155, 156] to alleviate OA symptoms. Although HA plays an important role in the joint, the clinical results of HA injections have been controversial;[155, 157-159] some clinical studies have shown benefits, but others have shown no statistical significant efficacy relative to placebo. We remark, however, that due to shear-thinning, the viscosity of such supplements (and synovial fluid itself) becomes comparable to that of water at the relevant shear rates at the cartilage surfaces. Directly-measured friction coefficients where a model surface is coated by HA reveal, however, that HA on its own is a rather poor boundary lubricant at physiological pressures (above a few atm), with $\mu \approx$ 0.2–0.5.[138, 142, 160] Benz et al.[160] found both surface adsorbed HA and covalently grafted HA resulted in friction coefficients between 0.15 and 0.3, far higher than the measured values for cartilage (e.g. **Table 1**). Thus HA on its own is not expected to be the molecule responsible for the remarkable lubricity of cartilage; however, it may contribute to wear protection of cartilage surfaces, shielding the underlying surfaces from damage even at contact stresses up to 200 atm during shear.[160, 161] Linking aggrecan molecules to HA attached to a model mica surface improves its boundary lubrication, with $\mu \approx$ 0.01, but only up to pressure of ca. 12 atm, increasing sharply at higher pressure, which suggests such complexes cannot account for the remarkable boundary lubrication observed in joints.[137, 162] More recently, Singh et al.[134] established a peptide-mediated binding of HA to cartilage, and the efficiency of HA as a boundary lubricant has been examined. Tissue surfaces treated with the HA-binding system exhibited good lubrication ($\mu \approx$ 0.008–0.014) up to a rather low pressure of 0.8 atm. *In vivo* experiments showed HA could be retained in the articular joint for a longer period of time through the HA-binding system compared with a control without HA-binding peptide.

Lubricin is an elongated (ca. 200 nm long, a few nm wide), mucinous glycoprotein (in human is encoded by the PRG4 gene), which is present both in the cartilage outer superficial zone and in synovial fluid,[163-165] whose properties have been extensively studied. It has been implicated in cartilage lubrication since this role was first suggested in 1970 by Radin and coworkers[87] (its somewhat suggestive name was coined in 1981 when it was identified as a mucin-like glycoprotein[166] which may be responsible for cartilage lubrication). Interestingly, this original study[87] showed that under conditions of their measurments the test lubricant (lubricin solution in the absence of HA) provided a rather moderate reduction in friction ($\mu$ = 0.0025) relative to saline ($\mu$ = 0.005), i.e. only a factor of 2. Researchers have reported that lubricin plays an important role in the boundary lubrication of cartilage,[53, 58, 117, 132, 167-171] and that its deficiency may be related to joint diseases, such as camptodactyly-arthropathy-coxa vara-pericarditis (CACP) syndrome,[169, 172] with the features of permanent bending of the fingers, synovial hyperplasia, and accelerated joint damage and failure; thus, its supplementation has been shown to delay disease progression.[58, 173]

Several direct measurements of friction between lubricin (or related molecules) boundary layers attached on model substrates have been carried out, using nanomechanical techniques such as SFB and friction-force microscopy (FFM), while more macroscopic techniques (macro-tribometry) used lubricin as an added lubricant for excised cartilage.[116, 117, 174] Zappone et al.[175] measured the normal and friction forces between layers of the human lubricin using an SFB, and showed that the friction coefficient between two hydrophilic mica surfaces bearing lubricin is relatively low ($\mu$ = 0.02–0.04) only under low pressures (up to ca. 6 atm) and increases to $\mu \approx 0.2$ under higher pressures. In a follow-up study[176] by the same group, the effect of enzymatic digestion by chymotrypsin of lubricin on the lubrication properties was studied. The chymotrypsin rapidly digests the end domains, releasing the nonadsorbing mucin domain from the surface. The anchoring of the glycosylated domain on the surface is thus critical for lubricin to express its full lubricating ability. A sharp or strong increase in the friction coefficient from $\mu = 0.02$–$0.04$ to $\mu = 0.13$–$1.17$ ocurrs when the pressure fell below 0.6 MPa (ca. 6 atm), attributed to non-lubricating ends of the lubricin. In a separate study, Harvey et al.[177] used the SFB to study the frictional properties of porcine gastric mucin, since, similar to lubricin, it is a glycoprotein with its dominant linear central region possessing the characteristic mucin structure. Low friction coefficients ($\mu \approx 0.02$–$0.03$) were seen up to mean pressures 7–10 atm, attributed to low interpenetration of the opposed layers together with hydration lubrication mechanism, with higher friction coefficients ($\mu \approx 0.1$–$0.2$) at higher pressure attributed to interlayer entanglements and to bridging, much as lubricin. In an FFM study, Chang et al.[178, 179] reported high friction coefficients ($\mu \approx 0.1$) were obtained by adding lubricin between two hydrophilic surfaces, much higher than what is expected for articular cartilage. More recently, and in contrast to the earlier studies, Huang et al.[180] found that low friction coefficients ($\mu \approx 10^{-2}$–$10^{-3}$) were obtained between recombinant human PRG4 (rhPRG4) on mica up to 3.6 MPa, but high friction coefficients ($\mu \approx 0.1$) when adding HA on top. It seems that boundary layers of natural lubricin,[175, 178] other mucins,[177] as well as layers consisting of lubricin with HA,[138, 140, 178, 179] with fibronectin together with lubricin,[140] and with collagen together with lubricin,[181] do not on their own provide lubrication at levels near that of articular cartilage at pressures higher than a few atmospheres. The precise mechanistic role of lubricin in articular cartilage lubrication, in the sense of knowing how its molecular structure contributes to low friction at the cartilage surface, is thus not yet fully elucidated. Other than working as a boundary lubricant, lubricin has been shown to play multiple functions in joints, including a role on chondroprotection by preventing synovial cell growth and protein adhesion on the cartilage surface.[59, 169, 182]

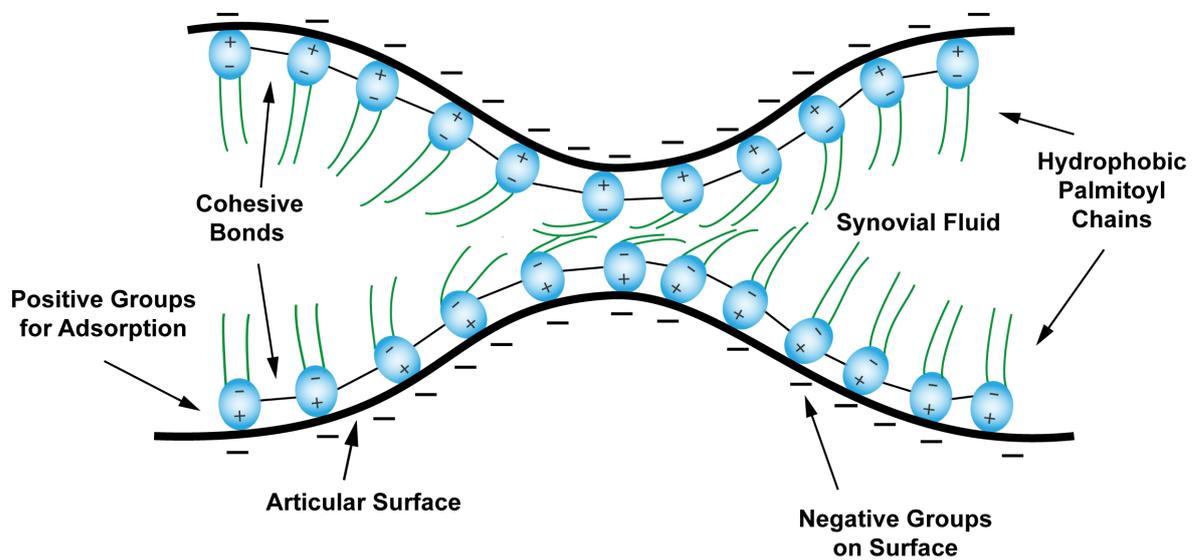

(A)

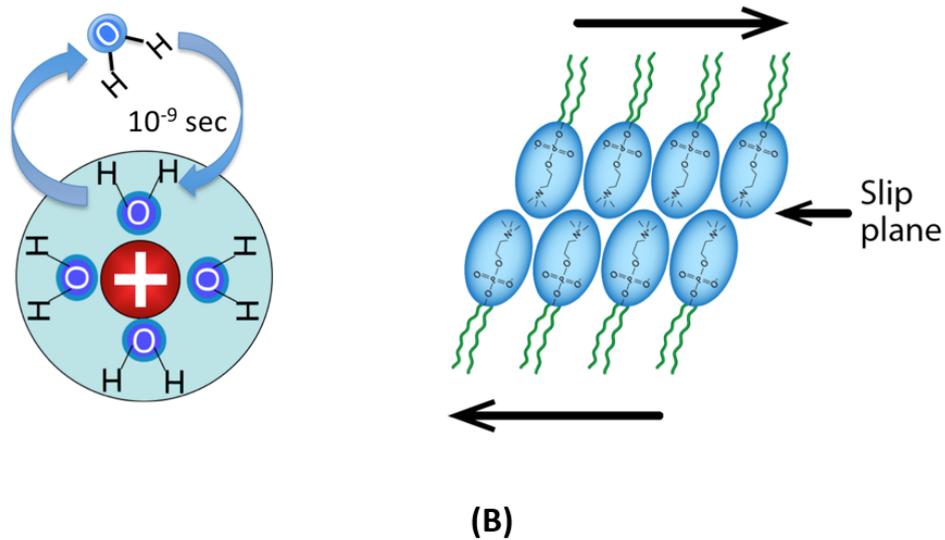

**(B)**

**Figure 5.** Proposed mechanisms for cartilage lubrication by PC lipids. (A) Lipid monolayers adsorb to the negatively-charged cartilage surface via their dipolar headgroups, exposing the di-acyl (fatty acid) chains to the interface, while frictional dissipation on sliding occurs as the chains slide past each other at the slip plane, much as in the classic boundary lubrication of metal surfaces by surfactants. Such a mechanism is known to lead to values of $\mu \approx 0.05$–$0.1$.[1] (B) In the hydration lubrication mechanism, assemblies of PCs (whether monolayer, bilayers or vesicles attached or complexed at the cartilage surface) exposing their phosphocholine headgroups slide past each other. Frictional dissipation takes place through shear of the hydration layers at the slip plane. This mechanism leads to low friction coefficients ($\mu \approx 0.001$ or less) up to high pressures (ca. 100 atm or more). Panel A reproduced with permission from ref. [20]. Copyright 2016, Annual Reviews. Panel B reproduced with permission from ref. [183]. Copyright 2015, American Chemical Society.

PCs, amphiphilic molecules consisting of two hydrophobic tails and a hydrophilic phosphocholine group, ubiquitous in synovial fluid and in cartilage as noted earlier, have also been proposed as the boundary lubricant of cartilage.[47, 48, 60, 128, 184] In 1969, Little et al.[97] loaded their joints up to 90 kg (maximal mean pressure estimated at ca. 10–20 atm). The friction coefficients did not vary with load or with sliding velocity, and over a measurement of 30 min period under high load the measured friction coefficients changed very gently only from 0.003 to 0.007. However, rinsing the articular surfaces with a good solvent (chloroform/methanol) for 'fats' (lipids), thereby presumably removing lipids or or reducing their concentration, increased friction coefficients by ~300%. Therefore, they strongly concluded the lubrication is boundary in nature, and the 'fat' (lipid) present in articular cartilage lowers the friction between cartilage surface. This work however received little attention at the time. Then, inspired by the high contact angle (100–103°) of water forming on the normal bovine cartilage surface after rinsing off synovial fluid with saline, Hills and Butler[185] proposed a model that surface-active phospholipids (SAPL) would adsorb via their polar (phosphocholine) headgroup on the cartilage surfaces (as illustrated in **Figure 5**A), exposing their hydrophobic acyl tails. They suggested that two such tail layers would slide past each other at the slip plane, much as boundary layers of surfactants, which are widely used for lubrication and wear reduction in machine tribology. Such a mechanism, however where frictional dissipation occurs via breaking and reforming of relatively-weak van der Waals bonds at the interface between the tails, is known to lead to friction coefficients $\mu \approx 0.05$–$0.1$,[1] much higher than for cartilage in joints, so that this mechanism would not appear to be correct.

Several earlier studies of lubrication by PC lipids, either on their own or together with HA, revealed that indeed friction coefficients in the range $\mu \approx 0.002$ to $> 0.5$ could be achieved,[97, 139, 185-189] the lower range comparable with that of healthy cartilage, depending on the PCs used. Trunfio-Sfarghiu et al.[188] found that 1,2-dipalmitoyl-sn-glycero-3-phosphocholine (DPPC) lipid bilayers in the solid phase could reduce friction to low values ($\mu \approx 0.002$, comparable to that found for cartilage) but only up to rather low pressures (3 atm). Utilizing an AFM colloidal probe technique, Wang et al.[190] found that DPPC bilayers could provide a low friction coefficient ($\mu < 0.03$) up to 42 MPa, which is close to 2 times higher than the load that the cartilage can sustain, and the load bearing capacity of the DPPC bilayer increases with increasing fluidity of the bilayer. In an SFB study, Yu et al.[191] found that adding dioleoylphosphatidylcholine (DOPC) lipids in the form of multilamellar vesicles (MLVs) to surfaces bearing HA in aqueous media didn't improve the boundary lubrication properties of HA ($\mu > 0.5$). Overall, however, no consistent mechanism which could account for the widely differing lubrication was identified in these earlier studies.

## 4. Hydration Lubrication

The studies described above reveal that the prime candidates for boundary lubrication, HA and lubricin, either by themselves or in combination, exhibit substantially larger boundary friction coefficients at physiological pressures than the low friction coefficients measured in whole joints. On the other hand, as noted above, the detailed frictional mechanism of PLs in the aqueous environment of the joint is not clear – the model of **Figure 5**A[185] clearly does not lead to the low friction of cartilage. The answer appears to lie elsewhere. In 2002, based on SFB measurements, it was proposed that the hydrated ions trapped between negatively charged mica in physiological ionic strength (ca. 0.1 M) may provide extremely efficient lubrication, through a mechanism termed 'hydration lubrication', acting as follows, and as illustrated on the left of **Figure 5B**.[192] Water molecules in the hydration shell surrounding an ion in an aqueous medium are generally very tenaciously attached to the enclosed charge, as a result of their large dipole (which results in reduction of the Born or self-energy of the charge). At the same time, the exchange of such hydration water molecules with surrounding free water molecules can be extremely rapid, with exchange rates up to $10^9$ s$^{-1}$ (depending on the nature of the enclosed charge). This rapid exchange rate implies comparably rapid relaxation of the hydration shell itself, and hence its high fluidity. Thus, the hydration shell surrounding a charge or ion between confining surfaces can strongly resist compression under load – that is, the water of hydration will not be squeezed out – while, when sheared, the compressed hydration layer behaves in a fluid-like manner. This combination of supporting a large compressive load, together with a fluid response to shear, results in striking lubrication properties in aqueous surroundings. In a recent SFB study, Ma et al.[10] identified the energy dissipation pathway and the viscous losses in the bound hydration layers alone for hydrated ions compressed between mica surfaces. They found an effective viscosity of the hydration shell which is ca. 250-fold larger than that of bulk water; such a value still enables very fluid-like response under shear. In a separate study, Donose et al.[193] observed a significant lubrication effect at high electrolyte concentrations between a silica particle and silica wafer, differing from the mica surfaces described above. More recently, Han et al.[194] reported that hydration lubrication also applies with trapped hydratedalkali metal ions between macroscopic silicon surfaces, achieving very low friction coefficients ($\mu \approx 10^{-3}$) measured by macroscopic tribometry up to 250 MPa.

The concept of hydration lubrication (unlike in the typically-oily environment of machines) provides a framework for understanding the origins of the very low boundary friction of cartilage ($\mu$ down to 0.001 or less) at physiological pressures, which has emerged over the past decade or so.[42, 183, 195-199] This mechanism was extended to several boundary lubrication systems in aqueous solution, including charged polymer,[200, 201] polyzwitterionic brushes,[202, 203] surfactants,[19, 204, 205] and in particular, liposomes[142, 206-209] or bilayers[41, 210, 211] of PC lipids, exposing phosphocholine groups at their outer surfaces. In the context of biological lubrication, the most interesting finding within the hydration lubrication paradigm is that zwitterionic phosphocholine groups, the headgroups of PC lipids, which are known to be highly hydrated with up to 15–20 or more water molecules in the primary hydration shell,[212-217] constitute extremely efficient lubrication elements ($\mu$ down to $10^{-4}$ or less at pressures up to over 200 atm). These values are consistent with the superior lubricity of living joints, were stable over extended sliding periods, and independent of sliding velocities over several orders of magnitude (indicating a boundary friction mechanism).

The lubrication itself between PC bilayer assemblies or PC-vesicle surfaces, takes place at the interface between the hydrated phosphocholine headgroup layers as they slide past each other, as illustrated in **Figure 5**B. Hydration lubrication thus provides the mechanism by which PC lipids may act to reduce friction at the cartilage boundary layer. Indeed, very recently an SFB study[218] of friction between boundary layers of PLs extracted from human and bovine synovial fluids and cartilage show them to provide similarly good lubrication ($\mu$ down to 0.001) up to high pressures (up to 100 atm or more). Radin et al.[86] studied the friction of bovine joints, loaded to some four times physiological load (500 kg). Even at such high loads (corresponding to ca. 100 atm stress between the cartilage surfaces), the friction coefficients were still as low as 0.011, and there was no difference between buffer and synovial fluid as the lubricant. They concluded that a hydrostatic mechanism was indicated because they could not conceive of a non-viscous squeeze film under such high pressures without some fluid hydrostatic pressure to bear up the load. Now however, we know that such 'squeeze films' can exist under such pressures, and that their viscosity is very low. We call them hydration layers – and we know they are held in place by the PC headgroups. These in turn are exposed by PC bilayers that are robust to normal stresses of 100 atm or more, a robustness arising from the strong interactions between the lipid tails.

**Table 2**. Summary of measured friction coefficients with different synovial joint boundary lubricants (or their analogs) in aqueous media on model surfaces, typical of studies noted in text. $v_s$ – sliding velocities. $P$ – Pressures.

| Boundary layers | Substrate | $\mu$ | Comments | Refs. |
|---|---|---|---|---|
| **HA based** | | | | |
| HA | Mica | ~ 0.2–0.5 | $P > 5$ atm; $v_s \approx 0.3$–3 µm/sec; SFB or SFA | [41, 137, 160, 191] |
| HA | Hydrophobized gold vs. hydrophilic or hydrophobized glass | ~ 0.1 | $P \approx 1$–2 atm; $v_s \approx 10$–40 µm/sec; FFM | [179] |
| HA | Poly(dimethylsiloxane) (PDMS) vs. silicon wafer | ~ 0.02–0.03 | $P \approx 0.8$ atm; $v_s \approx 0.1$ mm/sec; micro-tribometry | [219] |
| HA + aggrecan | Mica | ~ 0.01–0.1 | $P$ up to ca. 10 atm; $v_s \approx 0.3$ µm/sec; SFB | [137] |
| HA + fibronectin | Mica | ~ 0.4–1.1 | $P$ to > 3 atm; $v_s \approx 0.3$–30 µm/sec; SFA | [140] |
| **Lubricin (or mucin) based** | | | | |
| Lubricin | Mica | ~ 0.03–0.4 | $P$ up to ca. 6 atm; $v_s \approx 1$ µm/sec; SFA | [175] |
| Lubricin | Hydrophobized gold | ~ 0.07–0.6 | $P$ up to ca. 6 atm; $v_s \approx 1$ µm/sec; SFA | [175] |
| Lubricin | Polylysine | ~ 0.07–0.4 | $P$ up to ca. 6 atm; $v_s \approx 1$ µm/sec; SFA | [175] |
| Lubricin | Hydrophobized gold vs. hydrophilic or hydrophobized glass | ~ 0.1 | $P$ around 1 atm; $v_s \approx 10$–40 µm/sec; FFM; | [178, 179] |
| Porcine gastric mucin | Mica or hydrophobized mica | ~ 0.02–0.2 | $P$ up to ca. 7 atm; $v_s \approx 0.3$ µm/sec; SFB | [177] |
| Lubricin + HA | Hydrophobized gold vs. hydrophilic or hydrophobized glass | ~ 0.1 | $P \approx 1$–2 atm; $v_s \approx 10$–40 µm/sec; FFM | [178, 179] |
| Lubricin + HA | Mica | ~ 0.09–0.4 | $P$ up to ca. 40 atm; $v_s \approx 3$–100 µm/se; SFA | [138] |
| Lubricin + Type II collagen | Gold | ~ 0.1 | $P \approx 1$–2 atm; $v_s \approx 10$–40 µm/sec; FFM | [181] |
| Lubricin + COMP | Poly(methyl methacrylate) (PMMA) | ~ 0.06 | $P \approx 70$ atm; $v_s \approx 2$ µm/sec; FFM | [220] |
| Lubricin+ fibronectin | Mica | ~ 0.2–0.3 | $P > 40$ atm; $v_s \approx 0.3$–30 µm/sec; SFA | [140] |
| Lubricin + HA + Type II collagen | Gold vs. SiO$_2$ | ~ 0.01 | $P \approx 0.13$ atm; $v_s \approx 20$ µm/sec; FFM | [221] |
| **Phospholipid based** | | | | |
| 1,2-distearoyl-sn-glycero-3- | Mica | ~ $10^{-5}$–$10^{-4}$ | $P$ up to ca. 180 atm; $v_s \approx$ 0.5 µm/sec; SFB | [209] |
| phosphocholine (DSPC) | | | | |
| DPPC | Mica | ~ $10^{-4}$ | $P$ up to ca. 180 atm; $v_s \approx$ 0.5 μm/sec; SFB | [211] |
| DPPC | Silica | < 0.03 | $P$ up to ca. 420 atm; $v_s \approx$ 0.4 μm/sec; FFM | [190] |
| 1-palmitoyl-2-oleoyl-glycero-3-phosphocholine (POPC) | Mica | ~ $10^{-5}$ | $P$ up to ca. 160 atm; $v_s \approx$ 0.5 μm/sec; SFB | [211] |
| Egg sphingomyelin | Mica | ~ $10^{-4}$–$10^{-3}$ | $P$ up to ca. 100 atm; $v_s \approx$ 0.1-6 μm/sec; SFB | [222] |
| DOPC | Silica or hydrophobized silica | < 0.01 | $P$ up to ca. 30 atm; $v_s \approx$ 0.4–8 μm/sec; FFM | [187] |
| DPPC + HA | Quartz | ~ 0.004 | $P \approx$ 3 atm; $v_s \approx$ 1–10 cm/sec; macro-tribometry | [184] |
| DOPC + cross-linked HA | Mica | > 0.5 | $P$ up to ca. 20 atm; $v_s \approx$ 0.2–30 μm/sec; SFA | [191] |
| DPPC | Glass vs. hydroxy ethyl methacrylate (HEMA) hydrogel | ~ 0.002 | $P$ up to ca. 3 atm; $v_s \approx$ 0.1–1 mm/sec; macro-tribometry | [188] |
| DPPC + HA | Silica | ~ 0.02 | $P$ up to ca. 600 atm; $v_s \approx$ 2 μm/sec; FFM | [139] |

## 5. Synergies in Cartilage Lubrication

In the context of cartilage lubrication, PLs are clearly present on the cartilage surface and in the synovial fluid.[47, 60, 223, 224] Thus they may provide highly efficient lubrication, through the hydration lubrication mechanism enabled by the hydration layers surrounding the phosphocholine head groups of the lipids. The exact configuration of lipids on the cartilage surface still remains unclear, however. Using FFM, on lubrication by combined HA and PC vesicles, Wang et al.[139] found relatively low friction ($\mu \approx$ 0.01) between surfaces bearing layer-by-layer-deposited DPPC followed by HA, approaching the low friction coefficients found in synovial joints and having a high load bearing capacity (ca. 600 atm). A recent SFB investigation by Seror et al.[41] points to how PC lipids in joints may form suitable boundary layers at the cartilage surface. In this investigation, HA molecules were attached to a mica surface via avidin-biotin chemistry, to mimic their presence at the surface of articular cartilage (**Figure 6**). DPPC lipids (in their gel phase at the room temperature of the experiments) were then introduced in the form of small unilamellar vesicles (SUVs). PC lipids are known, from

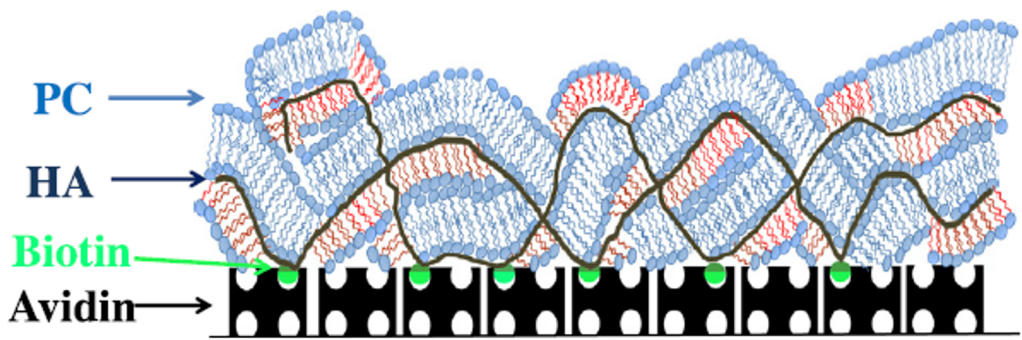

(A)

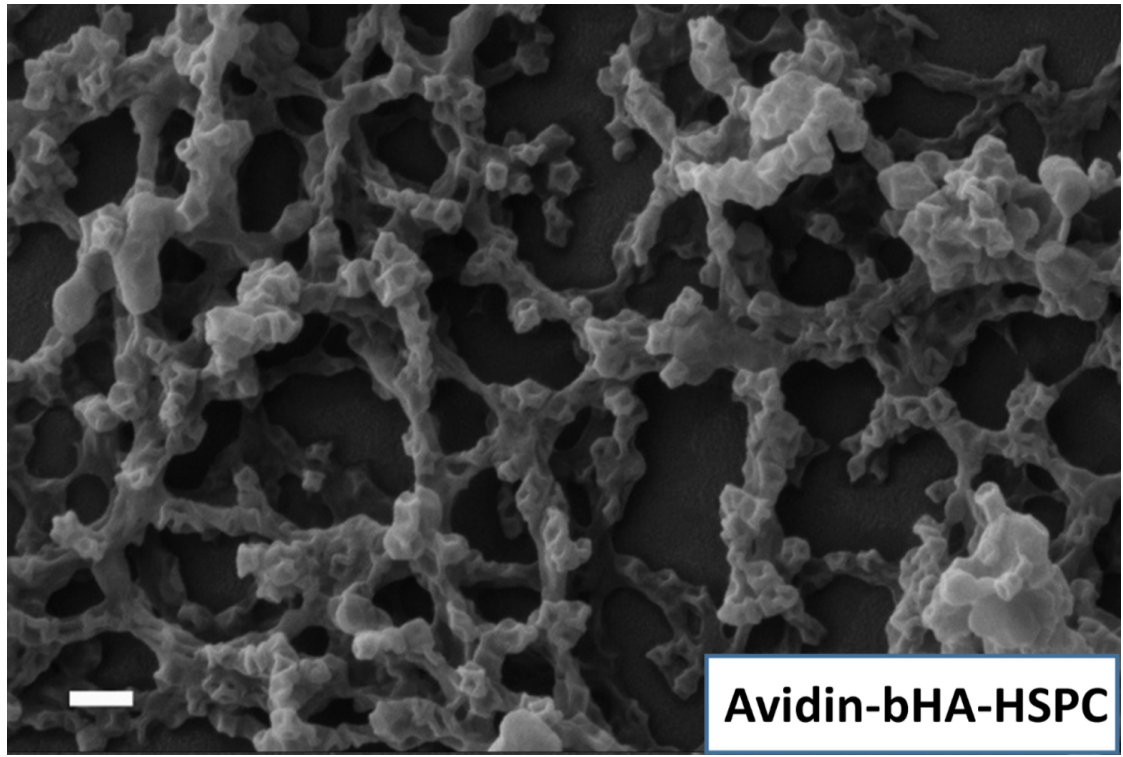

**(B)**

**Figure 6.** (A) Schematic of the HA/PC complexes on top of the avidin layer. The lipids may be in the form of bilayers (blue) attached to the negatively charged HA via a dipole-charge interaction with the dipolar phosphocholine PC headgroups (mostly), or monolayers (red) attached via the acyl chains of the PCs to the hydrophobic patches along the HA chain, or a combination of both, or indeed PC multilayers. [225-227] (B) Cryo-SEM image of freshly cleaved mica coated sequentially with avidin, biotinylated-HA and HSPC-SUVs, scale bars is 100 nm. Reproduced with permission from ref. [142]. Copyright 2017, Elsevier.

several different investigations, to interact with HA.[41, 122, 136, 226-229] This interaction may be either hydrophobic – between hydrocarbons on the phospholipid acyl chains and the hydrophobic patches on the HA molecules; or driven by dipole-charge interactions between the zwitterionic phosphocholine headgroups and the negatively-charged carboxylic group of HA, or a combination of the two.[225-227] More recently, Zhu et al.[142] extended this study to other PC lipids, mimicking the wide range of lipids in synovial joints; the resulting HA/PC surface complexes were imaged microscopically and characterized through surface forces (as shown in **Figure 6**). The friction measured for DPPC[41] and for hydrogenated soybean phosphatidylcholine (HSPC)[142] was very low ($\mu \approx 10^{-3}$) even at the pressures measured (ca. 150 atm), stable over extended sliding periods, and independent of sliding velocity over several orders of magnitude, characteristic of boundary lubrication. All these results led Seror et al.[41] to propose the following structure for the cartilage boundary layers and the mechanism of its lubrication: HA, ubiquitous in both cartilage and synovial fluid, is attached and exposed at the cartilage outer surface, where it complexes with PCs, which are likewise ubiquitous, to form robust boundary layers. These complexes are capable of providing the low friction ($\mu \approx 10^{-3}$) at the highest contact stress that have been measured in major joints (up to 100 atm or more),[21, 79] through the hydration lubrication mechanism. The attachment of the HA at the cartilage surface was suggested by Seror et al.[41] to occur via its known interactions with lubricin, which is known to be present both at the cartilage surface and in its outer superficial zone.[169, 221, 230, 231] Since HA is a long, linear flexible molecule, it may make multiple contacts with an adsorbing interface, so that even if the interaction at each contact is weak, such as van der Waals forces, charge-charge or charge-dipole interactions, the overall attachment of the HA to the lubricin molecules at the articular cartilage surface may be strong[232] (and may be further augmented by entanglement with the collagen or other microfibrillar network, such as fibronectin[140] or elastin[233] in the outer superficial zone). This proposed scenario is illustrated in the schematic in **Figure 7**, indicating how all three of the main synovial joint components act together, each with a very different role, to provide the lubrication characteristic of healthy joints. Clearly, since these three components must work together in the boundary lubrication of cartilage, this proposed scenario explains how a lubricin deficiency can affect joint friction[59, 87, 234] even when neither lubricin nor HA in themselves are the actual lubricating molecules, and also suggests possible new treatments to alleviate OA, for example intra-articular injections of PC lipids in the form of suitable vesicles with the addition of other molecules (HA or lubricin), which may augment the cartilage boundary lubricating layer. There is a potential shortcoming in this scenario, in that HA, which is abundant in the synovial fluid surrounding the cartilage, might be expected to adsorb and form highly dissipative bridges between the opposing, lipid-covered cartilage layers as they slide past each other, thereby leading to high friction as they slide. A very recent SFB study,[235] however, reveals directly that the presence of added HA to the intervening medium does not in fact affect the very efficient hydration lubrication seen in the absence of the added polysaccharide. The study is able to attribute this to the nature of the lipid-exposing boundary layer, together with the HA-PC interactions.

In a recent study, Raj et al.[220] created multilayer lipids using mixed HA/DPPC vesicle solution. These layers provide low friction coefficient ($\mu < 0.01$) with high load bearing capacity (ca. 20 MPa) and self-healing ability. Thus, one role of HA in synovial joints may be to enable lipid multilayers to accumulate on the cartilage surface and thus ensure a sufficient supply of these bio-lubricants at the sliding interfaces. This would fit the finding by Hills[184] that oligolamellar layers of PLs are present on the cartilage surfaces. Other than working as a carrier for lipid, HA molecules associated with lipids could inhibit their lysis by phospholipase A2.[236] Degraded HA molecules lose the capability of protecting lipids, which would lead to increasing friction. Thus HA plays an indirect role in the boundary lubrication process.

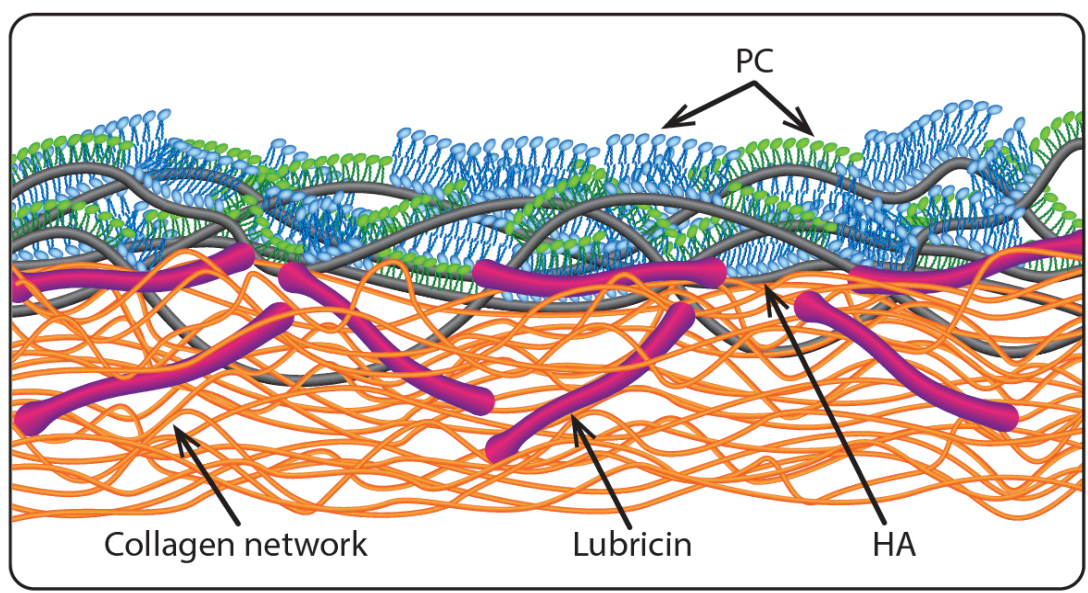

**Figure 7.** A schematic of the proposed structure of the boundary lubricating layer at the articular cartilage surface.[20] Lubricin molecules, known to be present both in the outer superficial zone and attached to the surface itself, interact with and immobilize HA molecules at the surface, which in turn complex with PC lipids (mostly via a dipole-charge interaction and partially via the hydrophobic interaction[225-227]), whose outer exposed highly-hydrated head-groups act to reduce friction via the hydration lubrication mechanism. Reproduced with permission from ref. [20]. Copyright 2016, Annual Review.

# 6 Natural, Synthetic and Bio-inspired Bio-lubricants

There is no known cure for OA, but treatments may reduce pain, improve movement, and slow down or prevent further cartilage degeneration;[237] since high joint friction is closely associated with tissue damage and thus OA, as described earlier, reducing such friction has been a major aim. Thus both naturally-occurring and synthetic molecules (themselves often inspired by nature) have been studied for their suitability for use as bio-lubricants to reduce cartilage friction and thereby alleviate OA. **Table 3** summarizes many of these investigations, while the following sections describe some of the main classes of lubricating molecules in more detail.

**Table 3**. Summary of measured friction coefficients with different macromolecular/supramolecular systems in aqueous media.

| System | Substrate | $\mu$ | Comments | Refs. |
|---|---|---|---|---|
| **Homopolymer** | | | | |
| Covalent attaching hyaluronic acid | Glass vs. gold | ~ 0.007; 0.18–0.33 | $P$ up to 10 atm; $v_s \approx 2$ μm/sec; FFM | [238] |
| Sodium poly(7-oxanorbornene-2-carboxylate) | Human cartilage | ~ 0.02 | $P \approx 1.3$ atm; $v_s \approx 0.3$ mm/sec; macro-tribometry | [239] |
| Chondroitin sulfate | Glass vs. bovine cartilage | ~ 0.05 | $P \approx 5$ atm; $v_s \approx 0.1$ mm/sec; macro-tribometry | [240] |
| Polyethylene glycol (PEG) | Mica | 0.01–0.05 | $P \approx 1$ atm; $v_s \approx 5$–1350 nm/sec; SFB | [241] |
| Poly(vinylpyrrolidone) (PVP) | Poly(dimethylsiloxane) (PDMS) vs. silicon wafer | 0.006–0.4 | Pressure $\approx 0.8$ atm; $v_s \approx 0.1$ mm/sec; micro-tribometry | [219] |
| **Copolymer** | | | | |

| Poly(methyl methacrylate)-block-poly(sodium sulphonated glycidyl methacrylate) (PMMA-b-PSGMA) | Hydrophobized mica | ≤ 10$^{-3}$ | $P$ up to ca. 3 atm; $v_s$ ≈ 0.3–3 µm/sec; SFB | [200] |
|---|---|---|---|---|
| poly(7-oxanorbornene-2-carboxylate) polymer containing pendent triethylene glycol | Bovine cartilage | ~ 0.03 | $P$ ≈ 7.8 atm; $v_s$ ≈ 0.3 mm/sec; macro-tribometry | [242] |
| Poly(L-lysine)-*graft*-poly(ethylene glycol) (PLL-*g*-PEG) | Mica | 0.001–0.003 | $P$ ≈ 250 atm; $v_s$ ≈ 0.5–5 µm/sec; SFA | [243] |
| Poly(L-lysine)-*graft*-Dextran (PLL-*g*-dex) | Glass | ~ 0.1 | $P$ ≈ 3400 atm; $v_s$ ≈ 1–19 mm mm/sec; macro-tribometry | [244] |
| Poly(methacryloxyethyl trimethyl-ammonium chloride)-b-poly(ethylene oxide)$_{45}$ ether methacrylate [(METAC)m-b-(PEO$_{45}$MA)n] | Silica vs. SiO$_2$ | ~ 0.03–0.04 | $P$ ≈ 500 atm; $v_s$ ≈ 2 µm/sec; FFM | [245] |
| Poly(quaternized poly[2-(dimethylamino)ethyl acrylate)-b-poly(ethylene oxide)$_9$ ether methacrylate (qPDMAEA$_{24}$-b-PEGMEA$_{400}$) | Mica or Bovine cartilage vs. glass | ~ 0.08–0.16 ~ 0.05–0.12 | $P$ ≈ 60 atm; $v_s$ ≈ 30 µm/sec; SFA $P$ ≈ 1.2 atm; $v_s$ ≈ 0.3 mm/sec; macro-tribometry | [120] |
| Catechol-anchored polymer | Mica | ~ 0.002–0.004 | $P$ up to ca. 20–30 atm; $v_s$ ≈ 2.5 µm/sec; SFA | [246] |
| Lubricin-mimics | Fibronectin anchored mica | ~ 0.28 | $P$ ≈ 34 atm; $v_s$ ≈ 0.3–30 µm/sec; SFA | [247] |
| Bioinspired bottle-brush polymer | Mica | ~ 0.001–0.01 | $P$ ≈ 20 atm; $v_s$ ≈ 0.01–20 µm/sec; SFA | [248] |
| Bottle-brush (BB) polymer (w/o HA) | Mica | | $v_s$ ≈ 3 µm/sec; SFA | [249] |
| a. mono-block BB | | ~ 0.01 | P up to ca. 4 atm | |
| mono-block BB with HA | | ~ 0.03 | High $\mu$ above ca. 15 atm | |
| b. di-block BB | | ~ 0.01 | High $\mu$ above ca. 15 atm | |
| di-block BB with HA | | ~ 0.01 | High $\mu$ above ca. 38 atm | |
| c. tri-block BB | | ~ 0.02 | High $\mu$ above ca. 80 atm | |
| tri-block BB with HA | | ~ 0.03 | $P$ up to ca. 140 atm | |
| Cartilage reactive graft copolymer | Bovine cartilage | ~ 0.01–0.08 | $P$ up to; $v_s$ ≈ 5 mm/sec; macro-tribometry | [250] |
| Cartilage reactive graft copolymer | PMMA vs. bovine cartilage | ~ 0.2–0.5 | $P$ up to; $v_s$ ≈ 5 mm/sec; macro-tribometry | [251] |
| **Surface-Initiated Polymer Brushes** | | | | |
| Poly[2-(methacryloyloxy) | Mica | ~ 10$^{-3}$– | $P$ ≈ 75–190 atm; $v_s$ ≈ 0.3 | [202, |
| -ethyl phosphorylcholine] (PMPC) | | $10^{-5}$ | µm/sec; SFB | [252] |
| PMPC | Silicon nitride vs. silicon | ~ 0.08 | $P \approx 1390$ atm; $v_s \approx 0.15$ m/sec; macro-tribometry | [253] |
| PMPC | Gold vs. silicon wafer | ~ 0.02–0.07 | Normal load to 9 nN; $v_s \approx 18$ µm/sec; FFM | [254] |
| PMPC | Cobalt–chromium–molybdenum alloy (Co-Cr-Mo) vs. polyethylene | ~ 0.01 | $P \approx 290$ atm; $v_s \approx 50$ mm/sec; macro-tribometry | [255] |
| Poly(oligo(ethylene glycol) methyl ether methacrylate) (POEGMA) | Co-Cr-Mo alloy vs. polyethylene | ~ 0.03 | $P \approx 290$ atm; $v_s \approx 260$ mm/sec; macro-tribometry | [255] |
| POEGMA | PDMS vs. Silica | ~ 0.02–0.04 | $P \approx 3$ atm; $v_s \approx 2$ mm/sec; macro-tribometry | [256] |
| PMETAC | PDMS vs. Silica | ~ 0.006 | $P \approx 2$ atm; $v_s \approx 2$ mm/sec; macro-tribometry | [257] |
| Poly(3-sulfopropyl methacrylate) potassium salt (PSPMA) | PDMS vs. Silica | ~ 0.005 | $P \approx 2$ atm; $v_s \approx 2$ mm/sec; macro-tribometry | [257] |
| Polystyrene sulfonate (PSS) a. 6 mM NaNO3 b. 0. 1 mM Y(NO3)3, 6 mM NaNO3 | Mica | ~ 0.005 ~ 0.15 | $v_s \approx 4.8$ µm/sec; SFA $P$ up to 50 atm $P$ up to 35 atm | [258] |
| **Lipid** | | | | |
| HSPC SUV | Mica | ~ $10^{-5}$–$10^{-4}$ | $P \approx 120$ atm; $v_s \approx 2$ µm/sec; SFB | [208] |
| 1,2-dimyristoyl-sn-glycero-3-phosphocholine (DMPC) MLV | Human cartilage | ~ 0.01–0.02 | $P \approx 24$ atm; $v_s \approx 0.5$–2 mm/sec; macro-tribometry | [259] |
| DPPC SUV | Ti6Al4V vs. polystyrene | ~ 0.007 | $P$ up to 150 atm; $v_s \approx 2.5$ µm/sec; FFM | [260] |
| HSPC SUV a. without HA b. with HA | Polyetheretherketone (PEEK) vs. silicon nitride | ~ 0.05 ~ 0.02 | $v_s \approx 0.8$–25.6 mm/sec; FFM $P \approx 127$ MPa $P \approx 127$ MPa | [261] |
| Stabilized HSPC SUV a. PEGylated b. PMPCylated | mica | 0.006–0.011 0.0006–0.001 | $v_s \approx 1$ µm/sec; SFB $P \approx 6$ MPa $P \approx 10$ MPa | [262] |
| **Others** | | | | |
| Polyelectrolyte brushes-grafted microgels | PDMS vs. silicon wafer | ~ 0.005–0.015 | $P \approx 4$ atm; $v_s \approx 1.6$ mm/sec; macro-tribometry | [263] |
| poly (3-sulfopropyl methacrylate potassium salt)-grafted mesoporous silica nanoparticles (MSNs-NH2@PSPMK) | Ti6Al4V vs. polyethylene (PE) | ~ 0.065 | $P \approx 26$–43.8 MPa; macro-tribometry | [264] |
| Polystyrene-poly(acrylic acid) core−shell particles | Amino-functionalized mica | ~ 0.04–0.3 | $P \approx 75$ atm; $v_s \approx 1.6$ µm/sec; SFB | [265] |
| Hydrophobin | Mica | ~ 0.15–1.4 | $P \approx 8$ atm; $v_s \approx 0.3$ μm/sec; SFB | [8] |
| Hydrophobin | Hydrophobized mica | ~ 0.14 | $P \approx 20$ atm; $v_s \approx 0.3$ μm/sec; SFB | [8] |
| Oligo(12-methacryloyldodecyl phosphorylcholine) (OMDPC) | Mica | ~ 0.001–0.006 | $P \approx 50$ atm; $v_s \approx 0.5$ μm/sec; SFB | [266] |

## 6.1 Homopolymers

HA is a naturally occurring nonsulfated glycosaminoglycan in aqueous solution, and believed to be important for the lubrication of joint cartilage.[133] Based on the unique mechanical properties of cellulose nanofibrils (CNF), which might mimic the mechanical environment of cartilage, HA was covalently bound to CNF and the tribological properties were tested by Valle-Delgado et al.[238] via an AFM colloidal probe technique. The friction coefficients ($\mu \approx 0.007$) obtained for HA attached CNF films were remarkable low at low ionic strength at a pressure of 10 atm, but increased to 0.15–0.33 at physiologically-high ionic strength, which might be due to weak hydration of HA molecules. As noted earlier, there is a controversy[267] surrounding the outcomes of intra-articular HA injections for OA, and one of the reasons is due to the enzymatic degradation and short residence time of HA in the joint.[134, 268-270] In order to bypass this shortcoming of HA, Wathier et al.[239] synthesized a non-biodegradable polyanion, sodium poly(7-oxanorbornene-2-carboxylate). This molecule reduced the friction between *ex vivo* human cartilages explants to $\mu = 0.06$, a lubricating effect superior to those of saline or Synvisc™ (a commercial hyaluronan formulation for intra-articular injection). This synthetic polyanionic lubricant is not readily degraded by hyaluronidase, and has no observed significant toxicity to human chondrocytes *in vitro*.

Chondroitin sulfate (CS) is a sulfated glycosaminoglycan, which is a major component of aggrecan, and which has become a widely used orally-taken supplement for alleviation of OA pain.[271] Its effect in patients with OA is possibly the result of the stimulation of the synthesis of proteoglycans and the decrease in catabolic activity of chondrocytes,[272] though a recent careful study suggested that orally-taken CS has an effect on OA alleviation comparable to that of placebo.[273] CS may however be potentially useful in reducing cartilage friction as an intra-articularly-injected additive. Thus Basalo et al.[240] investigated the effect of CS on the frictional response of bovine articular cartilage explants. They found that CS at high concentration significantly reduces the friction coefficient ($\mu = 0.05$) of bovine articular cartilage compared with that of control saline group ($\mu = 0.19$), neither mediated by viscosity nor osmolarity. Their results suggest that injection of CS into the joint cavity may have a potential therapeutic effect through the improving of cartilage tribological properties. Interestingly, Lawrence et al.[269] designed a lubricin-mimic (mLub) in which the backbone is chondroitin sulfate with covalently conjugated HA-binding peptides and type II collagen-binding peptides. Adding mLub reduces surface friction and adhesion of cartilage following trypsin treatment.

## 6.2 Polymer Brushes

Polymer brushes in good solvents have been widely studied as lubricating boundary layers.[202, 257, 258, 274-279] Their lubrication mechanism relies on the weak interpenetration of two opposing compressed brushes due to entropic factors (arising from excluded volume effects), even under high loads. Thus as two brushes slide past each other their interfacial region (where the slip plane lies) remains non-entangled and thus relatively fluid, resulting in low frictional dissipation on shear. This combination of sustaining high loads with a fluid interface results in low friction coefficients. In the context of joint lubrication, which is the theme of this perspective, polymer brushes may not only be exploited as possible boundary layers on articular cartilage, but may also be used to create lubricating boundary layers on the surfaces of artificial implants following joint replacement. Polymer brushes have also been studied in an attempt to mimic the structure of natural molecules such as lubricin. The first reports of polymer brush lubrication, by Klein et al.,[280] were carried out using polystyrene brushes in a good organic solvent (toluene), showing low friction coefficients ($\mu < 0.001$) between sliding surfaces up to contact pressures around 10 atm. This was later extended to charged or zwitterionic polymer brushes in aqueous media.[200, 202] There are two typical ways to create polymer brushes, both of which may be relevant to cartilage modification: 'grafting to' method and 'grafting from' method, in both of which it is necessary that the polymer does not attach or adsorb to the substrate other than by its end. The 'grafting to' method involves physical adsorption or chemical adsorption of pre-synthesized polymers that contain appropriate chain-end functional groups, which either have a high affinity to or can react with a complementary reactive group on the substrate of interest. Block-copolymers where one block adsorbs on the surface and the other is non-adsorbing have also been used to create polymer brushes.[120, 246, 248] The 'grafting-from' method is a bottom-up approach in which polymer chains are grown via surface-initiated polymerization from a substrate modified with functional groups that can initiate a polymerization reaction.[281-285]

Using a 'grafting to' method, Spencer and co-workers[243] have studied via macro and nanotribological approaches the lubrication properties of the polyelectrolyte graft copolymer poly(L-lysine)-graft-poly(ethylene glycol) (PLL-g-PEG), which is a water-soluble copolymer consisting of a poly(L-lysine) backbone and poly(ethylene glycol) side chains. The PLL chain carries multiple positive charges and spontaneously adsorbs onto negatively charged surfaces, such as many metal oxides or mica, exposing densely grafted PEG chains. In this context it is appropriate to note that the articular cartilage surface is negatively charged.[286] PEO is a neutral, biocompatible, chemically-inert, nonionic polymer that is utilized widely in both technological and biological applications due to its unusual solubility in both water and organic solvents. Measured in an SFB at pressures to ca. 100 atm, such brushes displayed a low friction coefficient, $\mu$ down to 0.001−0.003, in pure buffer solution at modest pressure (~30 atm). At higher pressures, film destruction was observed, immediately followed by a dramatic increase in the frictional force. By the addition of free graft-copolymer to the buffer solution, the resistance of the polymer brushes to compression and shear was strongly increased, an effect attributed to the self-healing capacity of the polymer boundary layer. The low friction of such adsorbed copolymer films render them potentially-useful as boundary layers for water-based lubrication. This self-healing capacity makes the polyelectrolyte-anchoring approach attractive for tribological applications in which the contacting surfaces are immersed in a reservoir of lubricant solution, such as synovial fluid surrounding the cartilage surfaces. It is of interest that at physiologically-high salt (of order 0.1 M salt concentration) the friction is higher than in pure water,[287] possibly due to competition by the salt ions for the hydration water. In contrast to uncharged PEG brushes, the monomers on charged or zwitterionic brushes may be much more extensively hydrated and thus able to reduce frictional dissipation strongly via the hydration lubrication effect. Raviv et al.[200] investigated the di-block copolymer (PMMA-b-PGMAS), with the hydrophobic PMMA moieties physically adsorbing to hydrophobized mica surfaces, while the nonadsorbing, negatively charged polyelectrolytic PGMAS tails extend into the aqueous medium. Up to pressures around 3 atm, the sliding friction between the charged brushes was extremely low, $\mu <$ ca. $6 \times 10^{-4}$, indicating very effective lubrication up to this pressure by the hydrated $-SO_3^-$ ions. At higher pressures, shearing the polymer-bearing surfaces led to a removal of the polymer from the surfaces, due to an increased friction between the hydrated charged monomers, which overcame the relatively weak hydrophobic polymer−surface adhesion. The increased friction is due to increased interpenetration and likely also to less efficient hydration lubrication, probably because the $-SO_3^-$ ion (in common also with anions more generally) is only weakly hydrated. Once the polymer is sheared off, the surfaces jump into adhesive contact and the friction coefficient increases sharply. In a recent SFB study,[248] a bottle-brush polymer showed quite low friction ($\mu \approx 3 \times 10^{-3}$). This polymer, inspired by the structure of bottle-brush glycoproteins (lubricin) (as shown in **Figure 8**), was prepared via the combination of atom transfer radical polymerization (ATRP) and post-modification techniques. The friction coefficients

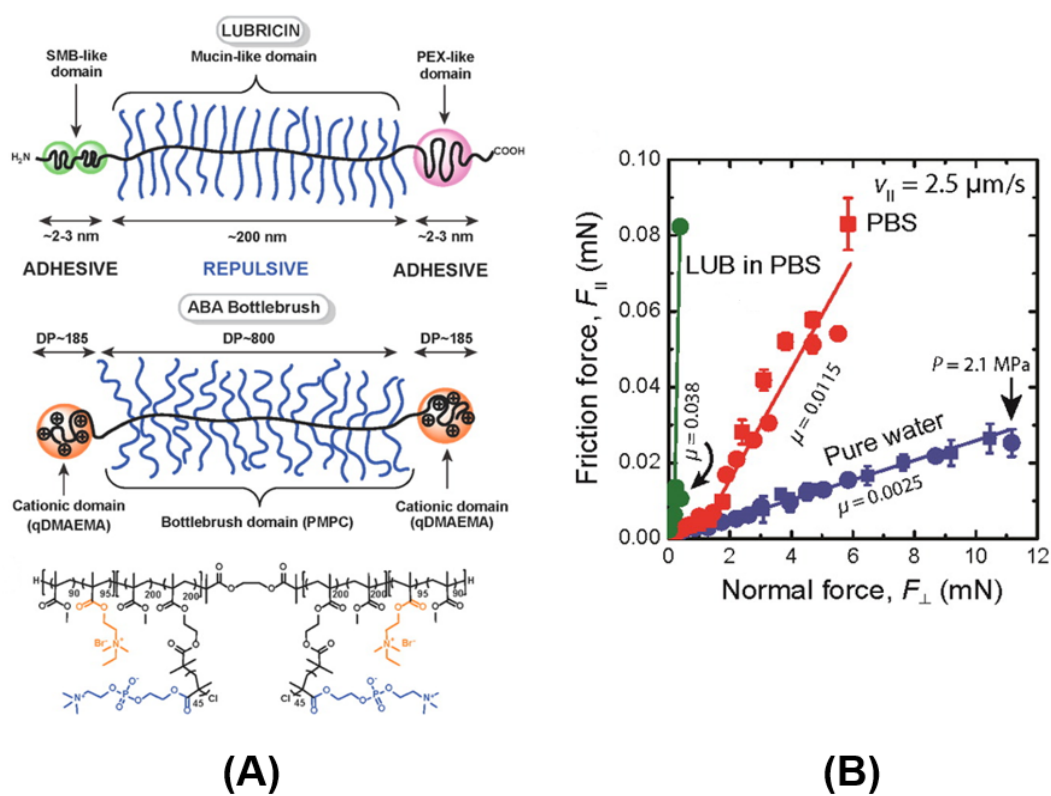

**Figure 8.** (A) Schematic representations of the protein lubricin (top) found in mammalian synovial fluids and the ABA bottle-brush polymer (bottom) mimicking lubricin. The bottle-brush polymer consists of two adhesive domains (cationic domains)

at its ends composed of a positively charged amine monomer (quaternized 2-(dimethylaminoethyl) methacrylate) (qDMAEMA) and a hydrophobic (methyl methacrylate) monomer, and a central bottle-brush domain, which is composed of a flexible backbone decorated with highly hydrated PMPC branches.[248] (B) The friction coefficients $\mu \approx 0.003$ between bottle-brush polymers are measured in pure water, while $\mu \approx 0.01$ in PBS solution. These values are below $\mu \approx 0.04$ reported for lubricin[175] under similar conditions. Reproduced with permission from ref. [248]. Copyright 2014, American Chemical Society.

between such bottle-brush boundary layers (ca. $10^{-3}$) is an order of magnitude higher than that between PMPC brush layers, likely due to the rougher nature of the bottle-brush polymer on the substrate to a uniform PMPC brush and the relatively lower density of the PMPC moieties on the bottle-brush structure (mushroom regime). For the development of OA treatments, Morgese et al.[250] developed graft-copolymers PGA-PMOXA-HBA with brush-forming, hydrophilic poly-2-methyl-2-oxazoline (PMOXA) side chains attached to a central backbone of polyglutamic acid and aldehyde-bearing tissue-reactive groups. Degraded cartilage treated with this graft-copolymer showed not only reduced protein adsorption, but also lower friction coefficients ($\mu \approx 0.01-0.08$) compared to native cartilages up to ca. 7 atm contact pressures (as shown in **Figure 9**). More recently, exploiting the ultra-dense brushes formed by cyclic loops on the surface,[288] Morgese et al.[289] further developed graft copolymers of cyclic PMOXA (instead of linear PMOXA), with a higher brush density and the same tissue-reactive groups. Such graft copolymers with cyclic side chains demonstrated better antifouling and lubrication properties on degraded cartilage compared to the linear-side-chain copolymers.[289] To reproduce the low friction coefficient and high pressure in joint cartilage, Faivre et al.[249] prepared a series of poly-phosphocholinated bottle-brush polymers (mono-block, di-block, tri-block) and investigated the tribological properties of these polymers with or without HA on mica. The friction coefficient of the bottle-brush polymers changed little ($\mu$ = 0.01-0.03) as the number of adhesive blocks on the bottle-brush polymer increased, but the critical pressure (above which the friction sharply rises, indicating breakdown of the lubricating boundary layer) increased drastically (from 0.4 MPa to 8.0 MPa). In the case of bottle-brush polymers with high molecular weight HA, the critical pressure increased significantly (~ 200–400%) compared to the bottle-brushes without HA. This synergy effect was also seen when PC lipids were mixed with HA,[41, 122, 142] which could be used to design new functional lubricants which are potentially useful for cartilage.

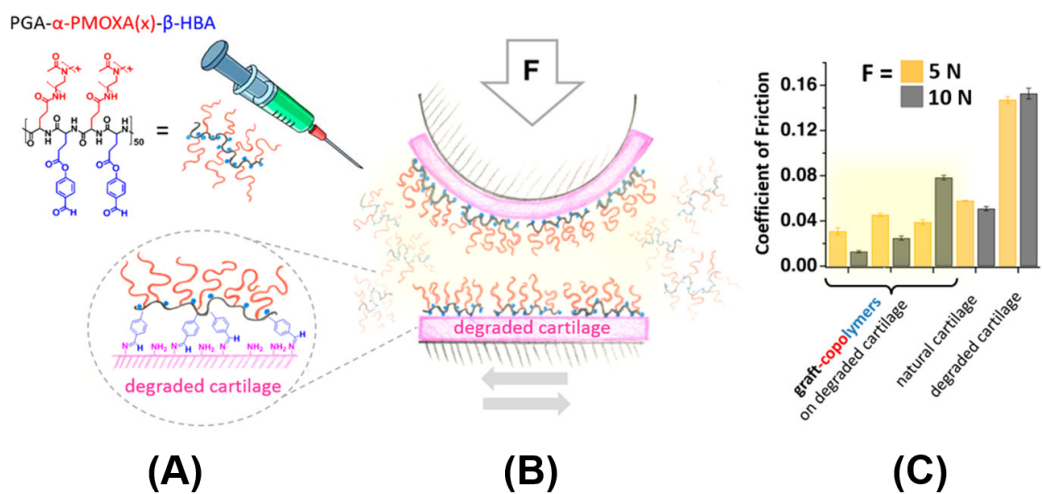

**Figure 9.** (A) Graft-copolymers PGA-PMOXA-HBA comprise a polyglutamic acid (PGA) backbone, coupled to brush-forming, PMOXA side chains, and to aldehyde-bearing tissue-reactive groups, for the anchoring on the degraded cartilage (the inset shows a schematic of immobilizing the graft-copolymer on degraded cartilage through Schiff-base linkages). (B) Degraded cartilage explants were coated with graft-copolymers by overnight incubation in PGA-PMOXA-HBA solutions. (C) The friction coefficients were measured by sliding them against each other immersed in a bovine synovial fluid solution of PGA-PMOXA-HBA at 5 mm/s and applying two different normal loads (5 N and 10 N). The graft-copolymer coatings reduced the friction coefficients of the degraded cartilage significantly under the tested conditions. Reproduced with permission from ref. [250]. Copyright 2017 American Chemical Society.

In the grafting-from method, higher grafting density and robust polymer brushes can be obtained. In the first SFB study of PMPC brushes, Chen et al.[202] reported that grafted-from PMPC brushes could lead to extremely efficient lubrication, with very low friction coefficients ($\mu$ = 0.0004) at pressures as high as 75 atm. This was attributed to the hydration lubrication mechanism, arising from the strong hydration of the phosphorylcholine groups on the PMPC chains. Recently, Tairy et al.[252] were able to graft even higher density PMPC brushes from mica surfaces, by reducing the level of dissolved oxygen in the system, and found even more efficient lubrication, with extremely low friction coefficients ($\mu$ = 7×10$^{-4}$–10$^{-5}$) at pressures as high as 150 atm. Thus these highly-hydrated poly-zwitterionic PMPC brushes were better at reducing friction better than neutral polymer or than polyanionic or polycationic brush layers.[255]

At the same time, PMPC brushes exhibited antifouling properties, and were able to prevent nonspecific protein adsorption onto surfaces, as well as to strongly reduce bacterial attachment and cell adhesion.[290-293] Since such fouling could change surface properties, thereby affecting tribological behavior, the grafting of such PMPC brushes with both antifouling and super-lubrication properties on joint implant surfaces seems very promising. Indeed, Moro et al.[294] grafted PMPC brushes onto the polyethylene acetabular of a hip prosthesis and observed greatly reduced wear (though the friction did not decrease as much, possibly due to asperity contact). Kyomoto et al.[255] investigated macroscopic tribological properties of differents hydrophilic polymer brushes grown from cross-linked polyethylene (CLPE) surfaces (as used also in joint prostheses) under high loads (up to 290 atm). The friction coefficients of zwitterionic PMPC brushes grafted onto CLPE ($\mu \approx 0.01$) was significantly lower than that of neutral POEGMA brushes ($\mu \approx 0.03$), polycationic brushes ($\mu \approx 0.14$), and polyanionic brushes ($\mu \approx 0.05$) grafted onto CLPE in simulated body fluid. This might be due to the higher level of hydration and the better capability of protein resistance in the grafted layer of the zwitterionic PMPC. We would note that most of the synthetic molecules described in this section, which may form efficient to extremely-efficient lubricating boundary layers, have potential for modifying synthetic surfaces – such as those of prosthetic devices – rather than on living cartilage.

## 6.3 Phospholipids

Outstanding lubrication performance by close-packed liposomes or lipid bilayers at physiologically high pressures was discovered using the SFB, a result which is particularly interesting since, unlike the synthetic molecules described in the previous section, lipids – particularly the PC type studied – are ubiquitous in biological systems. Goldberg et al.[208] created close-packed layers of small unilamellar vesicles (SUVs) of hydrogenated soy phosphatidylcholine (HSPC) on opposing mica surfaces. The friction coefficient between the liposome-coated mica surfaces was as low $\mu \approx 4 \times 10^{-4}$–$2 \times 10^{-5}$ at mean pressures up to 120 atm. The extreme lubricity of the liposome layers is attributed to the hydration lubrication mechanism acting via the strongly attached, fluid hydration layers surrounding the exposed phosphocholine groups at the outer liposome surfaces. Subsequently, Sorkin et al.[209] showed that the lubrication ability of liposome boundary layers was strongly correlated with the main (liquid-disordered to solid-ordered or gel phase) transition temperature $T_m$ of the corresponding lipid bilayers. When close-packed liposome layers slide past each other in lipid-free water (i.e. washed liposome layers), far better lubrication, and to higher contact stresses, was observed for lipids in their gel phase at the experimental temperature $T_{exp}$, i.e. when $T_m > T_{exp}$. This difference was attributed to the more robust nature of the gel-phase lipid bilayers. However, this trend is reversed when the measurements are carried out in dispersions of the liposomes. This unexpected effect was attributed to rapid self-healing of the liquid-phase bilayers, made possible by the presence of excess available lipids in the surrounding medium, together with the higher hydration level of the headgroups of the liquid-phase bilayers.[211] This effect may have implications for the understanding of complex boundary-layers present in the lubrication of joints, which contain a large variety of PC and other lipids, the majority of which, whether in synovial fluid or on the cartilage surfaces, are unsaturated lipids with low $T_m$ values.[47, 60, 295] Very recently Cao et al.[296] examined directly the effect of using lipid mixtures to create lubricating boundary layers. Sivan et al.[259] found that friction between cartilage explants in an aqueous dispersion of MLVs or of SUVs, composed of PC lipids, exhibited better lubrication properties ($\mu \approx 0.01$–$0.04$) relative to the aqueous medium alone ($\mu \approx 0.04$–$0.08$, no liposomes), with the lower friction values for the lower $T_m$ lipids in MLVs.

Whether one uses PC-SUVs or MLVs to introduce lipids to the fluid-immersed surfaces, including cartilage via intra-articular administration, it is essential, for any practical application, for the vesicles to maintain their stability with time towards aggregation. The gold-standard method for this is via steric stabilization of the vesicles by poly(ethylene glycol) chains that are incorporated in the lipid bilayer and stretch out from the surface (PEGylation). However, this may compromise the lubrication properties of the PC vesicles, due to the weak hydration of the PEG moieties in their corona.[208] By replacing the weakly-hydrated PEG chains that form the stabilizing layer around the vesicles by highly-hydrated PMPC moieties, Lin et al.[262] constructed

PMPCylated vesicles that show excellent stability against aggregation whether in water or aqueous salt solution. At the same time, such PMPCylated liposomes form boundary layers which provide a lubricity fully equivalent to that of unfunctionalized vesicles (recall however that the latter are unstable to aggregation), and which are an order of magnitude more lubricious than similar but PEGylated vesicles. Such PMPCylated liposomic vesicles may have widespread applicability, for example, for delivery of lubricity and thereby alleviation of widespread pathologies such as osteoarthritis (via intra-articular injection) or dry eye syndrome.[297-299] At the same time they may be used for lubrication of tissue engineering scaffolds for regenerative medicine, particularly in high stress environments (such as articular cartilage lesions) where delamination due to high shear stresses is a major cause of scaffold failure.[35] In this context we note that very recently,[300] inspired by the proposed boundary-lubrication mechanism of articular cartilage, hydrogels incorporating liposomes in their bulk exhibited sustained, very low friction due to lipid-based boundary layers exposed at their surface. This suggests that nature's lubrication solutions for cartilage may indeed be applied to synthetic gels.

## 7. Concluding Remarks and Perspective

We conclude this progress report by considering briefly the prospect of new materials and approaches based on the developing molecular-level understanding of biological lubrication processes, for a variety of advanced applications. The unique lubricity of healthy articular cartilage is important for its well-being, as increased friction at its surface may lead to a self-reinforcing cycle of wear and pro-catabolic chondrocyte gene-regulation, resulting in progressive tissue damage leading to OA. The presence of PC lipids in the boundary layer coating the cartilage, in structural synergy with HA and lubricin, and exposing their highly-hydrated phosphocholine groups at the slip plane,[20, 41, 301] has been proposed to provide such efficient lubrication within the emergent hydration lubrication paradigm.

Finally, we may identify a number of issues related to cartilage lubrication, whose elucidation would greatly advance our understanding, our ability to control such lubrication, and thereby, potentially, our ability to alleviate pathologies associated with high cartilage friction. These include:

1) The expression of mechanosensitive genes (including those responsible for cartilage degrading enzymes) in cartilage-embedded chondrocytes is regulated by biomechanical signals propagating from friction-induced shear stresses at the cartilage surface. It would be illuminating therefore to examine this directly *in vivo* by, for example, studying the effect of lipid-based lubricants on this regulation.

2) Maintenance of the integrity of the cartilage boundary layer, and thereby of cartilage homeostasis associated with its low friction, requires a better understanding of its structure and composition. This would open prospects for slowing or suppressing its degradation, for example through intra-articular administration of suitable PC formulations with HA and/or lubricin. Examining the behavior of this supramolecular boundary layer in model systems would directly lead to such improved insight.

3) Can liposomes could be designed to have dual functions, both as lubricants for articular cartilage and at the same time as versatile delivery vehicles which may deliver suitable drugs for treatment of cartilage pathologies?

4) Can self-lubricating surfaces be created, emulating cartilage lubrication by mimicking its boundary layer, which may be incorporated as low-friction, low wear-surfaces in joint prostheses? Could such surfaces be used also for tissue-engineering scaffolds in cartilage lesions, to eliminate scaffold-delamination in such high stress environments?

## Acknowledgements

The authors thank the McCutchen Foundation, the Rothschild Caesarea Foundation, the Israel Ministry of Science and Industry (grant 713272), and the Israel Science Foundation-National Science Foundation China joint program (grant ISF-NSFC 2577/17) for their support of this work and of work described in this review. This project has received funding from the European Research Council (ERC) under the European Union's Horizon 2020 research and innovation programme (grant agreement No. 743016). This work was made possible in part by the historic generosity of the Harold Perlman family.

Weifeng Lin received his Bachelor of pharmaceutical engineering in 2008, and his Ph.D degree in biochemical engineering in 2013 from Zhejiang University, China. Since 2013, he is a postdoctoral researcher in the Department of Materials and Interfaces, Weizmann Institute of Science, Israel. His research interests include hydration lubrication and biolubrication.

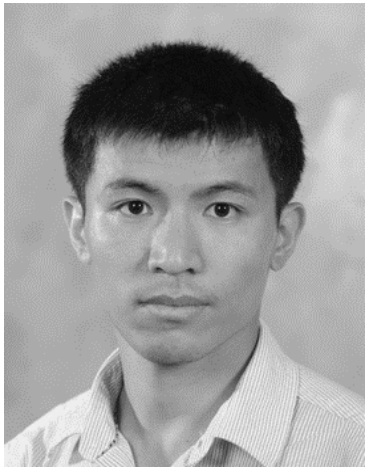

Jacob Klein is the Herman Mark Professor of Soft Matter Physics at the Weizmann Institute in Israel, where he has been since 1977. He was also on the faculty of Cambridge University's Physics Department (1980-1984), and from 2000-2007 he was the Dr. Lee's Professor of Chemistry at the University of Oxford and Head of its Physical and Theoretical Chemistry Department. His interests have ranged from the dynamics and interfacial properties of polymers to the behaviour of confined fluids and biological lubrication.

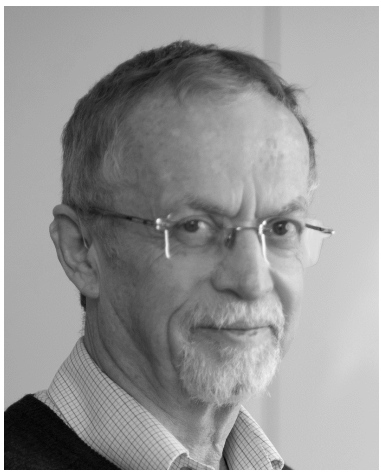